\documentclass[prb,twocolumn,showpacs]{revtex4}
\usepackage{graphicx} % \usepackage{bm} \usepackage{amsmath}

\usepackage{amsmath,amsthm,amssymb,mathrsfs}
\usepackage{bm}

\begin{document}

\title{Magnetization switching by current and microwaves}

\author{Tomohiro Taniguchi${}^{1}$, Daisuke Saida${}^{2}$, Yoshinobu Nakatani${}^{3}$, and Hitoshi Kubota${}^{1}$
      }
%\email{tomohiro-taniguchi@aist.go.jp}
 \affiliation{
  ${}^{1}$
  National Institute of Advanced Industrial Science and Technology (AIST), Spintronics Research Center, Tsukuba 305-8568, Japan \\
  ${}^{2}$
  Corporate Research and Development Center, Toshiba Corporation, Kawasaki 212-8582, Japan \\
  ${}^{3}$
  Graduate school of Informatics and Engineering, University of Electro-Communications, Chofu, Tokyo 182-8585, Japan %\\
%  ${}^{4}$
%  Graduate School of Engineering Science, Osaka University, Toyonaka, Osaka 560-8531, Japan
 }

\date{\today} 
\begin{abstract}
{
We propose a theoretical model of magnetization switching 
in a ferromagnetic multilayer by both electric current and microwaves. 
The electric current gives a spin transfer torque on the magnetization, 
while the microwaves induce a precession of the magnetization around the initial state. 
Based on numerical simulation of the Landau-Lifshitz-Gilbert (LLG) equation, 
it is found that the switching current is significantly reduced %by less than a half 
compared with the switching caused solely by the spin transfer torque 
when the microwave frequency is in a certain range. 
We develop a theory of switching from the LLG equation averaged over a constant energy curve. 
It was found that the switching current should be classified into four regions, 
depending on the values of the microwave frequency. 
Based on the analysis, we derive an analytical formula of the optimized frequency 
minimizing the switching current, 
which is smaller than the ferromagnetic resonance frequency. 
We also derive an analytical formula of the minimized switching current. 
Both the optimized frequency and the minimized switching current decrease with increasing the amplitude of the microwave field. 
The results will be useful to achieve high thermal stability and low switching current in spin torque systems simultaneously. 
}
\end{abstract}

 \maketitle

% ===================================================================================================================================================================================== %

\section{Introduction}
\label{sec:Introduction}

Magnetization switching in a ferromagnet has been an important topic in magnetism 
from the viewpoints of both fundamental physics and practical applications. 
The magnetization switching has been achieved by applying a direct magnetic field to a ferromagnet [\onlinecite{hubert98}]. 
In this case, the field magnitude should be larger than the anisotropy field $H_{\rm K}$, 
where, throughout this paper, we focus on the magnetization switching having an uniaxial anisotropy. 
Recent advances in fabrication of nanostructured ferromagnets with high-$H_{\rm K}$ materials however 
have required alternative methods for magnetization switching 
because of the technical difficulty of applying a large magnetic field to such a small ferromagnet. 
Spin transfer torque switching and microwave-assisted magnetization reversal 
are promising candidates as new methods for the magnetization switching. 
The spin transfer torque, or simply spin torque, is exerted on the magnetization 
by applying an electric current directly to a ferromagnetic multilayer 
consisting of a free and pinned layers [\onlinecite{slonczewski89,slonczewski96,berger96}]. 
The typical value of the switching current by the spin torque is on the order of $10^{6}-10^{7}$ A/cm${}^{2}$ 
[\onlinecite{katine00,grollier01,kiselev03,grollier03,krivorotov04,koch04,huai04,kubota05,kubota05IEEE,kiselev05,deac06,cui08,sukegawa10,yakushiji13,gopman14}]. 
On the other hand, in microwave-assisted magnetization reversal, 
microwaves with frequencies on the order of the ferromagnetic resonance (FMR) frequency 
efficiently supply energy to the ferromagnet, 
and reduce the direct field for switching. 
Typically, the switching field is reduced by less than half of the anisotropy field 
[\onlinecite{bertotti01,thirion03,denisov06,sun06,rivkin06,nozaki07,zhu08,okamoto08,okamoto10,okamoto12,tanaka13,barros11,barros13,cai13,klughertz15,taniguchi15APEX,kudo15}]. 

% ===================================================================================================================================================================================== %

The proposal of spin torque [\onlinecite{slonczewski89,slonczewski96,berger96}] drastically changed 
the theoretical view of magnetization switching from the switching due to a magnetic field. 
Let us assume that the magnetization dynamics can be described by the macrospin model. 
Then, the magnetization dynamics is regarded as a motion of a point particle in a potential having two energy minima. 
When a direct magnetic field is applied to the ferromagnet, 
the Zeeman energy modifies the shape of the potential. 
For a field magnitude larger than $H_{\rm K}$, the number of the minima is reduced to one; 
i.e., the potential has only one stable state. 
Then, the system finally moves to the stable state due to damping. 
This is the switching mechanism due to the magnetic field. 
On the other hand, the shape of the potential is unchanged in the spin torque switching. 
The spin torque competes with the damping torque. 
For a sufficiently large current, the spin torque overcomes the damping torque, 
and then the magnetization switches its direction by climbing up the potential landscape. 

% ===================================================================================================================================================================================== %

The microwave-assisted magnetization reversal provides an interesting theoretical example of the switching. 
It is convenient for understanding the switching mechanism to introduce a rotating frame [\onlinecite{bertotti09text}]. 
In this frame, the effect of the microwave is converted to an additional direct field 
pointing in the reversed direction. 
The additional field energetically stabilizes the switched state, and reduces the switching field. 
In this sense, the switching mechanism is similar to that due to a direct field. 
Simultaneously however, the microwave provides a torque preventing the switching [\onlinecite{taniguchi14PRB}]. 
Similarly to the spin torque, the direction of this torque is expressed by a triple vector product. 
Therefore, both the direct field effect and spin-torque-like effect coexist 
in the microwave-assisted magnetization reversal. 
Note however that the former assists the switching while the latter does not. 
Because of the competition between these two effects, 
the switching field in microwave-assisted magnetization reversal has a minimum at a certain microwave frequency; 
see Ref. [\onlinecite{taniguchi14PRB}] and Appendix \ref{sec:AppendixD}. 

% ===================================================================================================================================================================================== %

The spin torque switching has recently faced an unavoidable contradiction. 
It is desirable from the viewpoints of both fundamental and practical studies to 
reduce the spin torque switching current and enhance the thermal stability of the free layer simultaneously. 
Since the switching current is proportional to the anisotropy field $H_{\rm K}$ [\onlinecite{sun00,grollier03,morise05,taniguchi08PRB,suzuki09}], 
the reduction of the switching current can be achieved by using materials having relatively low $H_{\rm K}$. 
However, using low-$H_{\rm K}$ material leads to small thermal stability $\Delta_{0}=MH_{\rm K}V/(2 k_{\rm B}T)$, 
where $M$, $V$, and $T$ are the saturation magnetization, volume of the free layer, and temperature, respectively. 
Therefore, the reduction of the switching current and enhancement of the thermal stability is a trade-off problem. 
Here, note that the above discussion on microwave-assisted magnetization reversal provides a possibility to solve the problem. 
Let us consider the spin torque switching assisted by microwaves. 
We may find conditions of the microwave frequency to reduce the switching current, 
as in the case of the switching field in the microwave-assisted magnetization reversal. 
If the switching current is reduced under certain conditions, 
the method will be useful to satisfy both high thermal stability and low switching current simultaneously; 
i.e., the high thermal stability is guaranteed by using high-$H_{\rm K}$ material 
while the switching current can be reduced by applying microwaves. 
Numerical simulations have implied such a possibility [\onlinecite{carpentieri10,wang11JAP}]. 
However, the switching mechanism, as well as the theoretical conditions to reduce the switching current, has not been fully understood yet. 

% ===================================================================================================================================================================================== %

In this paper, we propose a theoretical model of a magnetization switching by current and microwaves. 
The electric current gives the spin torque on the magnetization, 
while the circularly polarized microwave induces the oscillation similarly to the case of the microwave-assisted magnetization reversal. 
Numerical simulation of the Landau-Lifshitz-Gilbert (LLG) equation reveals that 
the switching current for a certain microwave frequency is significantly reduced %by less than a half 
compared with the switching caused solely by the electric current. 
In fact, the switching current can become even zero when the microwave frequency is optimized 
and the amplitude of the microwaves becomes relatively high. 
Let us call the microwave frequency corresponding to the minimized switching currents the optimized frequency. 
To identify the optimized frequency, we develop a theory of the switching from the LLG equation averaged over a constant energy curve. 
It is found that the switching current should be classified into four regions, 
depending on the values of the microwave frequencies. 
Based on this analysis, we derive analytical formulas of the optimized frequency and the minimized switching current. 
It is found that the optimized frequency is smaller than the FMR frequency. 
Both the optimized frequency and the minimized switching current decrease with increasing the amplitude of the microwave field. 
The results indicate a possibility to simultaneously satisfy the requirements of high thermal stability and low switching current; 
i.e., using high-$H_{\rm K}$ material guarantees high thermal stability 
while the application of microwaves reduces the switching current. 

% ===================================================================================================================================================================================== %

The paper is organized as follows.
In Sec. \ref{sec:Numerical simulation}, 
we study the dependence of switching current 
on microwave frequencies by numerical simulation. 
It is found that the switching current is significantly reduced 
for a certain microwave frequency. 
In Sec. \ref{sec:Analytical theory}, 
we develop an analytical theory to explain the relation between the switching current and microwave frequency. 
The averaging method of the LLG equation over a constant energy curve is used 
to define the switching current analytically. 
The analytical solutions of an optimum frequency to reduce the switching current 
and minimized switching current are derived in Sec. \ref{sec:Optimized frequency and minimized switching current}. 
The conclusion is summarized in Sec. \ref{sec:Conclusion}. 

% ===================================================================================================================================================================================== %

% ===================================================================================================================================================================================== %

\begin{figure}%[p]
\centerline{\includegraphics[width=0.7\columnwidth]{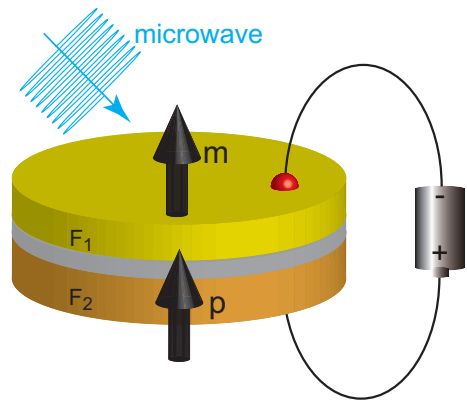}}%\vspace{-3.0ex}
\caption{
         Schematic view of the system. 
         The unit vectors pointing in the magnetization direction of the free and pinned layer are denoted as $\mathbf{m}$ and $\mathbf{p}$, respectively. 
         \vspace{-3ex}}
\label{fig:fig1}
\end{figure}

% ===================================================================================================================================================================================== %

% ===================================================================================================================================================================================== %

% ===================================================================================================================================================================================== %

\section{Numerical simulation}
\label{sec:Numerical simulation}

% ===================================================================================================================================================================================== %

\subsection{LLG equation}
\label{sec:LLG equation}

The system we consider is schematically shown in Fig. \ref{fig:fig1}. 
A ferromagnetic multilayer consists of the free and pinned layers 
separated by a nonmagnetic spacer. 
We assume that the magnetization dynamics in the free layer is described by the macrospin model. 
The unit vectors pointing in the magnetization direction 
of the free and pinned layers are denoted as $\mathbf{m}$ and $\mathbf{p}$, respectively. 
We assume that both the free and pinned layers are perpendicularly magnetized, and $\mathbf{p}=\mathbf{e}_{z}$. 
The current directly flowing in the multilayer gives the spin torque 
on the magnetization $\mathbf{m}$, 
where the positive current corresponds to 
the electrons flowing from the free to pinned layer. 
A circularly polarized microwave with a constant frequency $f$ is 
applied to the free layer. 
The magnetization dynamics in the free layer is described by the LLG equation given by 
\begin{equation}
  \frac{d \mathbf{m}}{dt}
  =
  -\gamma
  \mathbf{m}
  \times
  \mathbf{H}
  -
  \gamma 
  H_{\rm s}
  \mathbf{m}
  \times
  \left(
    \mathbf{p}
    \times
    \mathbf{m}
  \right)
  +
  \alpha
  \mathbf{m}
  \times
  \frac{d \mathbf{m}}{dt},
  \label{eq:LLG}
\end{equation}
where $\gamma$ and $\alpha$ are 
the gyromagnetic ratio and the Gilbert damping constant, respectively. 
The magnetic field, $\mathbf{H}$, acting on the free layer consists of 
the uniaxial anisotropy field along the $z$ axis 
and the circularly polarized microwave field rotating in the $xy$ plane. 
The explicit form of the magnetic field is given by 
\begin{equation}
  \mathbf{H}
  =
  \begin{pmatrix}
    H_{\rm ac} \cos 2\pi ft \\
    H_{\rm ac} \sin 2\pi ft \\
    H_{\rm K} m_{z}
  \end{pmatrix},
  \label{eq:H_field}
\end{equation}
where $H_{\rm ac}$ and $H_{\rm K}$ are 
the magnitude of the microwave field and the uniaxial anisotropy field, respectively. 
Note that the magnetic field is related to the magnetic energy density $E$ via 
\begin{equation}
\begin{split}
  E
  &=
  -M 
  \int d \mathbf{m}
  \cdot
  \mathbf{H}
\\
  &=
  -MH_{\rm ac}
  \left[
    m_{x} 
    \cos(2\pi ft)
    +
    m_{y}
    \sin(2\pi ft)
  \right]
  -
  \frac{MH_{\rm K}}{2}
  m_{z}^{2}.
  \label{eq:H_energy}
\end{split}
\end{equation}
The magnetization dynamics can be regarded as 
a motion of a point particle in an energy landscape of $E$, as mentioned above. 
Since the magnetic field $\mathbf{H}$ explicitly depends on time, 
the energy landscape of $E$ also changes in time. 
For analytical theory, it is convenient to introduce a rotating frame, 
as discussed in Sec. \ref{ref:Transfer to rotating frame}, where the energy landscape is independent of time. 
The ferromagnet has two stable states, $\mathbf{m}=\pm\mathbf{e}_{z}$, in the absence of the microwaves. 
The first term on the right-hand side of Eq. (\ref{eq:LLG}) induces a precession of the magnetization around the magnetic field. 
The precession direction is a counterclockwise (clockwise) rotation 
when the initial state of the magnetization is $\mathbf{m}=+(-)\mathbf{e}_{z}$. 
On the other hand, the rotation direction of the microwave field is 
counterclockwise (clockwise) for positive (negative) frequency $f$. 
The spin torque strength $H_{\rm s}$ is given by 
\begin{equation}
  H_{\rm s}
  =
  \frac{\hbar \eta j}{2eMd},
  \label{eq:H_s}
\end{equation}
where $\eta$, $j$, and $d$ are 
the spin polarization of the current, 
the current density, 
and the thickness of the free layer, respectively. 
In the following, 
we use the following parameter values obtained from typical experiments 
on spin torque switching and microwave-assisted magnetization reversal: 
$M=1000$ emu/c.c., 
$H_{\rm K}=7.5$ kOe, 
$H_{\rm ac}=450$ Oe, 
$\gamma=1.764 \times 10^{7}$ rad/(Oe$\cdot$s), 
$\alpha=0.01$, 
$\eta=0.6$, 
and $d=2$ nm. 
Note that the FMR frequency 
of the free layer is $f_{\rm FMR}=\gamma H_{\rm K}/(2\pi) \simeq 21$ GHz. 

% ===================================================================================================================================================================================== %

% ===================================================================================================================================================================================== %

\begin{figure}%[p]
\centerline{\includegraphics[width=1.0\columnwidth]{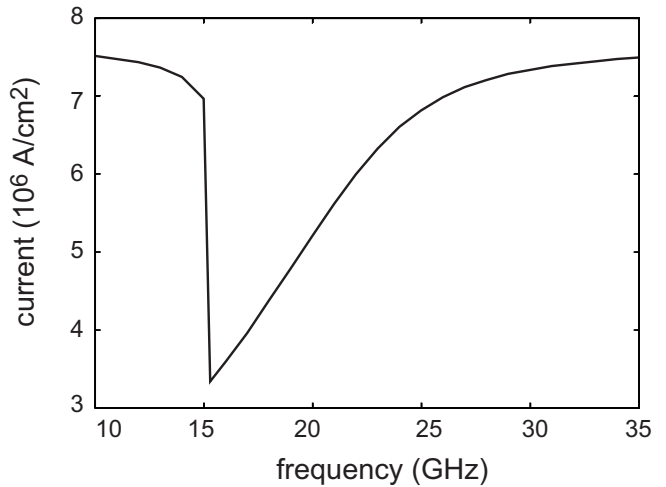}}%\vspace{-3.0ex}
\caption{
         Numerically evaluated switching current as a function of microwave frequency. 
         \vspace{-3ex}}
\label{fig:fig2}
\end{figure}

% ===================================================================================================================================================================================== %

% ===================================================================================================================================================================================== %

\subsection{Numerically evaluated switching current}
\label{sec:Numerically evaluated switching current}

We estimate the switching current by solving Eq. (\ref{eq:LLG}) numerically. 
In our simulation, the initial state of the magnetization is chosen as $\mathbf{m}(0)=+\mathbf{e}_{z}$, for convention. 
The switching current is defined as a minimum current necessary for satisfying $m_{z}(t_{\rm max})<-0.9$, 
where the current and microwaves are applied from $t=0$ to $t=t_{\rm max}$. 
Note that the switching current with this definition depends on the value of $t_{\rm max}$. 
The switching current becomes large for a short $t_{\rm max}$ 
because a large spin torque is necessary to achieve a fast switching. 
The switching current decreases with increasing $t_{\rm max}$, 
and saturates to a certain value. 
We numerically confirmed that $t_{\rm max}=100$ ns is sufficient 
to obtain the saturated lowest value of the switching current using our parameters. 
Therefore, $t_{\rm max}=100$ ns is adopted in the following simulation. 

% ===================================================================================================================================================================================== %

Figure \ref{fig:fig2} shows the dependence of the switching current 
on the microwave frequency $f$. 
Starting from a low value of $f$, 
the switching current decreases with the increase of the frequency. 
Around $f \sim 15-16$ GHz, 
the switching current discontinuously drops to the lowest value. 
The optimized frequency in our parameters is $15.3$ GHz, 
at which the switching current is about $3.3 \times 10^{6}$ A/cm${}^{2}$ 
(see also Fig. \ref{fig:fig11} below). 
Above that, the switching current increases with the increase of the frequency. 

% ===================================================================================================================================================================================== %

\subsection{Discussion}
\label{sec:Discussion}

We note that the switching currents shown in Fig. \ref{fig:fig2} 
in the limits of $f \to 0$ and $f \to \infty$ saturate to 
the critical current derived for the switching caused solely by the spin torque, 
\begin{equation}
  j_{\rm c}
  =
  \frac{2 \alpha eMd}{\hbar \eta}
  H_{\rm K},
  \label{eq:jc_spin_torque}
\end{equation}
which is $7.6 \times 10^{6}$ A/cm${}^{2}$ for our parameters. 
On the other hand, 
the switching current is about $3.3 \times 10^{6}$ A/cm${}^{2}$ 
at the optimized frequency, as mentioned above. 
Therefore, the switching current is reduced by less than half of Eq. (\ref{eq:jc_spin_torque}) 
by applying the microwaves. 
This fact indicates that the proposed switching model will be useful 
to realize high thermal stability and low switching current simultaneously in spin torque switching; 
i.e., using high-$H_{\rm K}$ material guarantees the high thermal stability, 
while applying microwaves reduces the switching current [\onlinecite{patent}]. 

% ===================================================================================================================================================================================== %

One might be interested in an experimental situation to test the above numerical result. 
A candidate is to put a ferromagnetic multilayer consisting of 
a giant magnetoresistance (GMR) or magnetic tunnel junction (MTJ) device on a coplanar waveguide. 
Another candidate is a GMR/MTJ device directly connected to a spin torque oscillator (STO). 
Such system was recently realized in experiment [\onlinecite{suto14}]. 
In the system, an electric current injected directly into the multilayer excites spin torque on the magnetization in the free layer of the GMR/MTJ device. 
Simultaneously, the current induces a self-oscillation in the STO. 
Then, the STO emits a dipole field to the free layer, which plays a similar role to microwaves. 
Therefore, it will be possible to test the present proposal experimentally. 
In such system, the microwave frequency becomes time dependent 
because the microwave originates from the dynamic coupling between the free layer and the STO through the dipole interaction. 
The magnetization dynamics by microwaves having time-dependent frequency is an attractive topic 
in the field of microwave-assisted magnetization reversal [\onlinecite{rivkin06,barros11,barros13,cai13,klughertz15,taniguchi15APEX,kudo15}]. 
For simplicity however, we focus on a constant frequency only in this paper. 

%In practice, a microwave source is necessary. 
%In particular, in practical application such as magnetic memory, 
%the size of the microwave source should be comparable or smaller than the size of the target ferromagnet 
%to avoid an undesirable switching in a different ferromagnet in integrated memory. 
%A candidate of such microwave source is a spin torque oscillator (STO). 
%Recently, microwave assisted magnetization reversal excited by microwaves 
%generated by an STO directly connected to a ferromagnetic multilayer was experimentally observed \cite{suto14}. 
%Using similar system, it will be possible to test the present proposal experimentally. 
%In such system, however, the microwave frequency becomes time dependent 
%because of the dipole coupling between the target ferromagnet and the STO. 
%Thus, a further improvement of the present model will be necessary, 
%which is beyond the scope of this paper. 

% ===================================================================================================================================================================================== %

Another candidate to satisfy high thermal stability and low switching current simultaneously 
is voltage control of magnetic anisotropy [\onlinecite{ohno00,chiba03,weisheit07,chiba08,maruyama09,shiota09,endo10,nozaki10,chiba11,wang11,shiota11,nozaki12,pertsev13}]. 
Reduction of the perpendicular anisotropy, $H_{\rm K}$, by the electric field 
results in the reduction of the switching current. 
A combination of the spin torque effect and the voltage control of the magnetic anisotropy will be 
another interesting subject for future magnetic recording applications. 
%In this case, however, relatively thick oxide is necessary to induce the voltage effect, 
%which might cause an increase of the resistance or restriction in choosing materials. 

% ===================================================================================================================================================================================== %

An important question in the present results is to identify the relation between the optimized frequency, the minimized switching current, and material parameters. 
In the next section, we show a detailed analysis of the magnetization switching to answer this question. 

% ===================================================================================================================================================================================== %

% ===================================================================================================================================================================================== %

\section{Analytical theory}
\label{sec:Analytical theory}

The purpose of this section is to clarify the relation between 
the optimized frequency, the minimized switching current, and the material parameters. 
To this end, we develop an analytical approach to explain 
the relation between the microwave frequency and the switching current. 

% ===================================================================================================================================================================================== %

% ===================================================================================================================================================================================== %

% ===================================================================================================================================================================================== %

\begin{figure}%[p]
\centerline{\includegraphics[width=1.0\columnwidth]{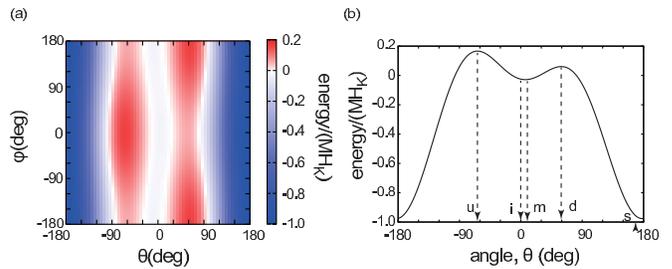}}%\vspace{-3.0ex}
\caption{
         (a) An example of the energy landscape of $\mathscr{E}$ with $f=10$ GHz. 
             The angles, $\theta$ and $\varphi$, are defined as $\theta=\cos^{-1}m_{z^{\prime}}$ and $\varphi=\tan^{-1}(m_{x^{\prime}}/m_{y^{\prime}})$, respectively. 
             The value of $\mathscr{E}$ is normalized by $MH_{\rm K}$. 
         (b) The potential $\mathscr{E}$ along the line $\varphi=0$. 
             The symbols, i, m, d, u, and s mean the initial state, metastable state, saddle point, maximum (unstable) point, and stable (switched) state, respectively. 
         \vspace{-3ex}}
\label{fig:fig3}
\end{figure}

% ===================================================================================================================================================================================== %

% ===================================================================================================================================================================================== %

\subsection{Transfer to rotating frame}
\label{ref:Transfer to rotating frame}

To develop an analytical theory, 
it is convenient to transfer from the laboratory frame 
to a rotating frame $x^{\prime}y^{\prime}z^{\prime}$, 
in which the $z^{\prime}$ axis is parallel to the $z$ axis 
and the $x^{\prime}$ axis is always pointing to the direction of the rotating field. 
The LLG equation in the rotating frame is given by [\onlinecite{bertotti09text}] 
\begin{equation}
\begin{split}
  \frac{d \mathbf{m}^{\prime}}{dt}
  =&
  -\gamma 
  \mathbf{m}^{\prime}
  \times
  \bm{\mathcal{B}}
  -
  \left(
    \gamma 
    H_{\rm s}
    -
    \alpha 
    2 \pi f 
  \right)
  \mathbf{m}^{\prime}
  \times
  \left(
    \mathbf{e}_{z^{\prime}}
    \times
    \mathbf{m}^{\prime}
  \right)
\\
  &
  -\alpha
  \gamma
  \mathbf{m}^{\prime}
  \times
  \left(
    \mathbf{m}^{\prime}
    \times
    \bm{\mathcal{B}}
  \right),
  \label{eq:LLG_rotating_frame}
\end{split}
\end{equation}
where $\mathbf{m}^{\prime}=(m_{x^{\prime}},m_{y^{\prime}},m_{z^{\prime}})$ is 
the unit vector pointing in the direction of the magnetization in the rotating frame. 
We neglect higher order terms of $\alpha$ because the Gilbert damping constant is small ($\alpha \ll 1$) in typical ferromagnets [\onlinecite{oogane06}]. 
The magnetic field in the rotating frame is given by 
\begin{equation}
  \bm{\mathcal{B}}
  =
  \begin{pmatrix}
    H_{\rm ac} \\
    0 \\
    -(2\pi f/\gamma) + H_{\rm K} m_{z^{\prime}}
  \end{pmatrix}.
  \label{eq:B_field}
\end{equation}
As in the case of the laboratory frame, 
we define the energy density in the rotating frame as 
\begin{equation}
\begin{split}
  \mathscr{E}
  &=
  -M \int 
  d \mathbf{m}^{\prime}
  \cdot
  \bm{\mathcal{B}}
\\
  &=
  - M H_{\rm ac}
  m_{x^{\prime}}
  +
  M \frac{2\pi f}{\gamma}
  m_{z^{\prime}}
  -
  \frac{MH_{\rm K}}{2}
  m_{z^{\prime}}^{2}.
  \label{eq:B_energy}
\end{split}
\end{equation}
The first term on the right-hand side of Eq. (\ref{eq:B_energy}) is 
the Zeeman energy with the microwave field $H_{\rm ac}$, 
which points to the positive $x^{\prime}$ direction in the rotating frame. 
The second term indicates that a magnetic field, $(2\pi f/\gamma)$, pointing in 
the negative (positive) $z^{\prime}$ direction appears in the rotating frame for positive (negative) microwave frequency, 
as pointed out in Ref. [\onlinecite{bertotti09text}]. 
This field makes the switched state energetically stable 
when the magnetization and microwaves rotate in the same direction. 
The last term is the uniaxial anisotropy energy. 

% ===================================================================================================================================================================================== %

Contrary to the laboratory frame, where the energy density is given by Eq. (\ref{eq:H_energy}), 
Eq. (\ref{eq:B_energy}) does not explicitly depend on time. 
Therefore, the magnetization dynamics can be regarded as a motion of a point particle in a fixed landscape. 
This is an advantage to use the rotating frame. 
We emphasize that high symmetry of the present model along the $z$-axis enables us to use the frame. 
For example, if the cross section of the free layer is elliptic 
or the microwaves are linearly polarized [\onlinecite{carpentieri10,wang11JAP}], 
we cannot introduce such a frame. 
While the scope of this paper is the magnetization switching of a perpendicular ferromagnet, 
excitation of the periodic or quasiperiodic mode in a spin torque oscillator by microwaves 
was studied by using the rotating frame [\onlinecite{bonin09,dAquino11}]. 

% ===================================================================================================================================================================================== %

Figure \ref{fig:fig3}(a) shows an example of the potential $\mathscr{E}$ for $f=10$ GHz, 
where two angles, $\theta$ and $\varphi$, are defined as 
$\theta=\cos^{-1}m_{z^{\prime}}$ and $\varphi=\tan^{-1}(m_{x^{\prime}}/m_{y^{\prime}})$, respectively. 
The magnetization dynamics can be regarded as 
a motion of a point particle in such energy landscape. 
Although the zenith angle is usually defined in the range of $0^{\circ} \le \theta \le 180^{\circ}$, 
it is convenient to use the range of $-180^{\circ} \le \theta \le 180^{\circ}$ in this figure 
because it clarifies the location of the maximum point of $\mathscr{E}$. 
The potential $\mathscr{E}$ is symmetric with respect to the $x^{\prime}z^{\prime}$ plane at which $\varphi=0$, 
and the energetically stable states, saddle point, and unstable state exist in this plane. 
Therefore, in the following, 
we frequently use the cross section of the energy landscape along the line $\varphi=0$. 
Figure \ref{fig:fig3}(b) is an example of such figure; 
i.e., it is $\mathscr{E}$ along the line $\varphi=0$ in Fig. \ref{fig:fig3}(a). 
The symbols i, m, d, u, and s mean the initial state, metastable state, saddle point, maximum (unstable) point, and stable (switched) state, respectively. 
Since the microwave field, $H_{\rm ac}$, points to the positive $x^{\prime}$ direction in the rotating frame, 
the metastable and stable states locate in the region $m_{x^{\prime}}>0$, 
while the maximum point appears in the region $m_{x^{\prime}}<0$. 
Note that the initial state, $\theta(t=0)=0$, has higher energy than the metastable state. 
This is because the Zeeman energy with the microwave field in Eq. (\ref{eq:B_energy}), which appears from $t=0$, 
modifies the energy landscapes between $t<0$ and $t>0$; 
i.e., before the application of the microwaves ($t<0$), the perpendicular state, $\theta=0$, is stable, 
while after turning off the microwaves from $t=0$, the stable state shifts to a position of $m_{x^{\prime}}>0$. 

% ===================================================================================================================================================================================== %

\subsection{Competition between spin torque and microwaves}
\label{sec:Competition between spin torque and microwaves}

As mentioned above, the effect of microwaves on the magnetization dynamics in the rotating frame 
is to add an additional field, $(2\pi f/\gamma$), along the $z^{\prime}$ axis. 
Note that the microwaves give another term given by $\alpha 2\pi f \mathbf{m}^{\prime} \times (\mathbf{e}_{z^{\prime}} \times \mathbf{m}^{\prime})$ 
in Eq. (\ref{eq:LLG_rotating_frame}). 
The fact that the direction of this torque is expressed by a triple vector product, similarly to the spin torque, 
indicates that the mathematical analysis for spin torque driven magnetization dynamics 
can be applied to the analysis of microwave-assisted magnetization reversal or vice versa [\onlinecite{bertotti09text,taniguchi14PRB,suto15}]. 
It should be emphasized that this torque moves the magnetization to the initial equilibrium state 
when the microwaves rotate in the same direction with the magnetization precession; 
i.e., it prevents the switching. 
This is because the precession is stabilized by such microwaves, 
and therefore, the magnetization stays near the equilibrium state. 
As discussed in Ref. [\onlinecite{taniguchi14PRB}], 
because of this term, a finite magnetic field is necessary for microwave-assisted magnetization reversal 
even when the potential $\mathscr{E}$ has only one minimum. 
%this term causes a jump of the switching field in microwave assisted magnetization reversal. 
The term plays a similar role in the present model, 
as will be mentioned in Sec. \ref{sec:Classification of switching current}. 
One might consider that the microwaves rotating in the opposite direction to the precession assist the switching 
because the spin-torque-like term by such microwaves points to the switched state. 
Usually, such microwaves are inefficient for switching because the switched state becomes energetically unstable. 
It was recently shown, however, that such microwaves result in the switching when the frequency depends on time [\onlinecite{taniguchi15APEXa}]. 

% ===================================================================================================================================================================================== %

We emphasize that the application of microwaves is useful to reduce the switching current,
even though the spin-torque-like term prevents the switching 
because the switching current becomes less than half of Eq. (\ref{eq:jc_spin_torque}), as shown in Fig. \ref{fig:fig2}. 
Moreover, in Fig. \ref{fig:fig12} below, we show that 
the switching current can be zero with the optimized frequency 
when the amplitude of the microwaves becomes relatively high. 

% ===================================================================================================================================================================================== %

% ===================================================================================================================================================================================== %

\begin{figure}%[p]
\centerline{\includegraphics[width=1.0\columnwidth]{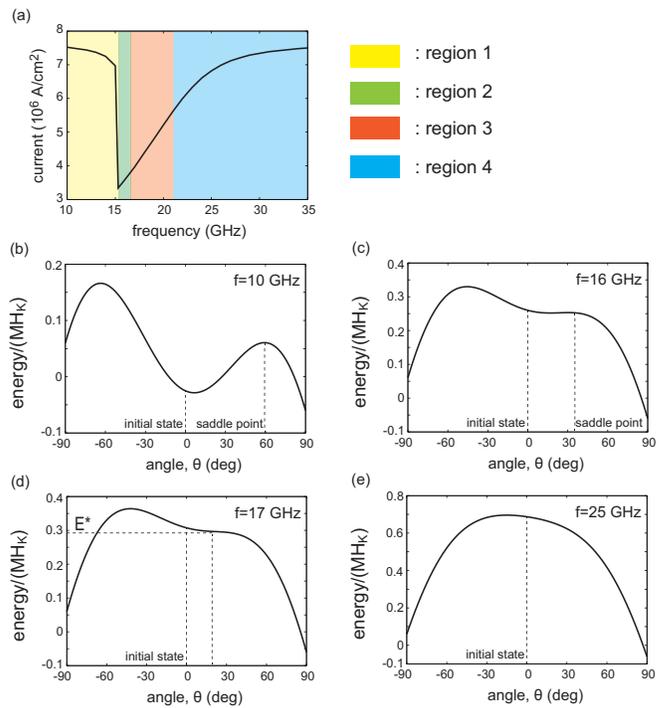}}%\vspace{-3.0ex}
\caption{
         (a) The classification of the switching current into the four frequency regions. 
         Typical energy landscapes of regions 1, 2, 3, and 4 are shown in 
         (b) $f=10$ GHz, (c) $f=16$ GHz, (d) $f=17$ GHz, and (e) $f=25$ GHz, respectively. 
         \vspace{-3ex}}
\label{fig:fig4}
\end{figure}

% ===================================================================================================================================================================================== %

% ===================================================================================================================================================================================== %

\subsection{Classification of switching current}
\label{sec:Classification of switching current}

As analyzed in the following discussion, 
it is necessary to classify the switching current into four regions, 
depending on the microwave frequencies. 
This is because the microwave in the rotating frame gives a field along the $z$ direction, 
as can be seen in Eq. (\ref{eq:B_energy}), 
and changes the shape of the energy landscape of $\mathscr{E}$. 
Figure \ref{fig:fig4}(a) illustrates this classification. 
Here, we summarize the relation between the microwave frequency and the shape of the energy landscape 
in each region. 

\textit{Region 1}. 
      This region is defined by the microwave frequency less than the optimized frequency. 
      Figure \ref{fig:fig4}(b) shows a typical energy landscape in region 1. 
      The energy landscape has a saddle point, 
      and the initial state of the magnetization is energetically lower than the saddle point. 

\textit{Region 2}. 
      This region is defined by the microwave frequency around $f \sim 16$ GHz. 
      A typical energy landscape in region 2 is shown in Fig. \ref{fig:fig4}(c). 
      The energy landscape has a saddle point, 
      and the magnetization initially locates in higher energetic state than the saddle point. 

\textit{Region 3}. 
      This region is defined by the microwave frequency of $17 \lesssim f \lesssim 21$ GHz. 
      As shown in Fig. \ref{fig:fig4}(d), 
      the energy landscape does not have a saddle point. 
      Instead, there is a point corresponding to an energetic condition $\mathscr{E}=\mathscr{E}^{*}$, 
      at which the second derivative of $\mathscr{E}$ becomes zero. 
      The reason why a finite current is necessary to switch the magnetization, 
      in spite of the fact that the potential barrier has only one minimum, 
      is that the second term on the right-hand side of Eq. (\ref{eq:LLG_rotating_frame}), 
      $\alpha 2\pi f \mathbf{m}^{\prime} \times (\mathbf{e}_{z^{\prime}} \times \mathbf{m}^{\prime})$, 
      prevents the switching, 
      as in the case of microwave-assisted magnetization reversal [\onlinecite{taniguchi14PRB}]. 

\textit{Region 4}. 
      This region is defined by the microwave frequency of $f \gtrsim 22$ GHz. 
      The gradient of $\mathscr{E}$ becomes monotonic from the unstable to the stable state. 
      We note that such change of the potential shape between region 3 and region 4 occurs 
      at the FMR frequency, $f_{\rm FMR}=\gamma H_{\rm K}/(2\pi)$. 

% ===================================================================================================================================================================================== %

% ===================================================================================================================================================================================== %

% ===================================================================================================================================================================================== %

% ===================================================================================================================================================================================== %

\begin{figure}%[p]
\centerline{\includegraphics[width=1.0\columnwidth]{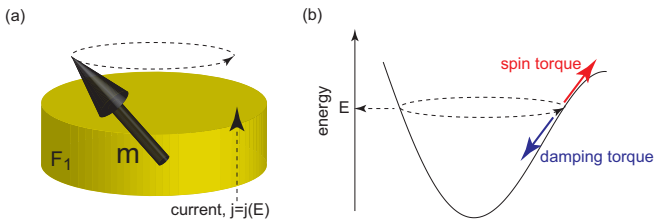}}%\vspace{-3.0ex}
\caption{
         (a) Schematic representation of a steady precession of the magnetization in the rotating frame 
         and (b) corresponding trajectory displayed on the energy landscape. 
         \vspace{-3ex}}
\label{fig:fig5}
\end{figure}

% ===================================================================================================================================================================================== %

% ===================================================================================================================================================================================== %

\subsection{Balance current}
\label{sec:Balance current}

The switching occurs as a result of the competition between the spin and damping torques, 
which correspond to the second and third terms 
on the right-hand side of Eq. (\ref{eq:LLG_rotating_frame}). 
Therefore, it is useful to define the balance current 
at which the damping torque balances with the spin torque. 
Since the spin and damping torques are balanced, 
the magnetization dynamics is mainly determined by 
the field torque, $-\gamma \mathbf{m}^{\prime} \times \bm{\mathcal{B}}$ in Eq. (\ref{eq:LLG_rotating_frame}). 
Thus, the magnetization precesses many time during the switching, as shown below. 
Note that the precession occurs on a constant energy curve of the energy landscape of $\mathscr{E}$ because the field torque conserves the energy. 
Thus, strictly speaking, the balance current is defined 
from the spin and damping torques averaged over a precession trajectory 
on a constant energy curve of $\mathscr{E}$. 
The physical picture of the balance current is schematically shown in Fig. \ref{fig:fig5}. 
In Fig. \ref{fig:fig5}(a), 
the magnetization shows a steady precession in real space. 
The current $j$ is equal to the value of a balance current $j(\mathscr{E})$. 
This precession corresponds to a rotating motion of a point particle 
on a constant energy curve of $\mathscr{E}$, as shown in Fig. \ref{fig:fig5}(b), 
where the spin torque cancels the damping torque. 
The balance current, or the LLG equation averaged over a constant energy curve, 
has been used to derive theoretical conditions of 
spin torque switching, spin-torque-induced self-oscillation, thermally activated magnetization switching, 
and microwave-assisted magnetization reversal 
[\onlinecite{bertotti04,bertotti05,bertotti07,bertotti09text,serpico04,serpico05,serpico06,hillebrands06,bazaliy11,bazaliy11JAP,dykman12,newhall13,taniguchi13PRB1,taniguchi13PRB2,taniguchi14PRB,lacoste13,taniguchi13,pinna13,pinna14,taniguchi15,suto15}]. 

% ===================================================================================================================================================================================== %

The balance current is estimated by the following equation, 
\begin{equation}
  \oint 
  d t 
  \frac{d \mathscr{E}}{dt}
  =
  0,
  \label{eq:dEdt_zero}
\end{equation}
where the integral range is over one precession period 
on a constant energy curve of $\mathscr{E}$. 
According to Eqs. (\ref{eq:B_field}) and (\ref{eq:B_energy}), 
Eq. (\ref{eq:dEdt_zero}) can be rewritten as 
\begin{equation}
  \mathscr{W}_{\rm s}(\mathscr{E})
  +
  \mathscr{W}_{\alpha}(\mathscr{E})
  =
  0,
  \label{eq:dEdt_zero_another}
\end{equation}
where $\mathscr{W}_{\rm s}$ and $\mathscr{W}_{\alpha}$ are given by 
\begin{equation}
  \mathscr{W}_{\rm s}
  =
  \oint 
  dt 
  M 
  \left(
    \gamma 
    H_{\rm s}
    -
    \alpha 
    2\pi f
  \right)
  \left[
    \bm{\mathcal{B}}
    \cdot
    \mathbf{e}_{z^{\prime}}
    -
    \left(
      \mathbf{m}^{\prime}
      \cdot
      \mathbf{e}_{z^{\prime}}
    \right)
    \left(
      \mathbf{m}^{\prime}
      \cdot
      \bm{\mathcal{B}}
    \right)
  \right],
  \label{eq:W_s}
\end{equation}
\begin{equation}
  \mathscr{W}_{\alpha}
  =
  -\oint 
  dt 
  \alpha
  \gamma
  M
  \left[
    \bm{\mathcal{B}}^{2}
    -
    \left(
      \mathbf{m}^{\prime}
      \cdot
      \bm{\mathcal{B}}
    \right)^{2}
  \right].
  \label{eq:W_alpha}
\end{equation}
Equation (\ref{eq:W_s}) is the change of energy during a precession on a constant energy curve of $\mathscr{E}$ 
due to the second term on the right-hand side of Eq. (\ref{eq:LLG_rotating_frame}). 
This term can be either positive or negative, 
depending on the parameters such as $f$ and $j$. 
On the other hand, Eq. (\ref{eq:W_alpha}) is the dissipation due to damping, and is always negative 
because this term arises from the damping torque in Eq. (\ref{eq:LLG_rotating_frame}). 
From Eq. (\ref{eq:dEdt_zero_another}), 
the balance current $j(\mathscr{E})$ is expressed as 
\begin{equation}
  j(\mathscr{E})
  =
  \frac{2 \alpha eMd}{\hbar \eta}
  \left\{
    \frac{2\pi f}{\gamma}
    +
    \frac{\oint d t [\bm{\mathcal{B}}^{2}-(\mathbf{m}^{\prime}\cdot\bm{\mathcal{B}})^{2}]}
      {\oint dt [\bm{\mathcal{B}}\cdot\mathbf{e}_{z^{\prime}} - (\mathbf{m}^{\prime}\cdot\mathbf{e}_{z^{\prime}})(\mathbf{m}^{\prime}\cdot\bm{\mathcal{B}})]}
  \right\}.
  \label{eq:jE}
\end{equation}

% ===================================================================================================================================================================================== %

As mentioned above and shown in Fig. \ref{fig:fig5}(b), 
when the current $j$ equals a value $j(\mathscr{E})$, 
the magnetization precesses on a constant energy curve of $\mathscr{E}$ 
because the spin torque cancels the damping torque, 
and therefore, only the field torque, $-\gamma \mathbf{m}^{\prime} \times \bm{\mathcal{B}}$, 
exciting the precession remains finite. 
On the other hand, when the current $j$ is larger than any values of $j(\mathscr{E})$, 
the magnetization cannot stay on any constant energy curves of $\mathscr{E}$, 
and moves along the gradient of the energy landscape to the switched state. 
Thus, the switching current, $j_{\rm sw}$, can be defined from the balance current as 
\begin{equation}
  j_{\rm sw}
  =
  {\rm max}
  [j(\mathscr{E})]. 
  \label{eq:jsw_def}
\end{equation}
The range of $\mathscr{E}$ in Eq. (\ref{eq:jsw_def}) 
should be discussed for each region, as shown below. 
The use of the balance current to estimate the switching current is applicable for small damping $(\alpha \ll 1)$ due to the following reason. 
The derivation of $j(\mathscr{E})$ assumes that 
the averaged work done by spin torque and dissipation due to damping 
during a precession on a constant energy curve balance each other. 
At each point on the constant energy curve however, the magnitudes of the spin torque and the damping torque are not equal 
because these torques have different angular dependencies. 
For a large $\alpha$, a shift from the constant energy curve at a certain point becomes large, 
and the magnetization cannot return to the constant energy curve during a precession. 
Then, the assumption in the averaged LLG equation does not stand, 
and the balance current is no longer applicable to estimate the switching current. 
The appropriate range of $\alpha$ can be, in principle, estimated 
by comparing the energy change during a precession and 
the energy difference between the minimum (initial) and maximum (or saddle) point [\onlinecite{taniguchi13PRB1}]. 

% ===================================================================================================================================================================================== %

We note that the balance current and 
the critical current, $j_{\rm c}$, in spin torque switching relate via $j_{\rm c}=\lim_{\mathscr{E} \to \mathscr{E}_{\rm min}, f \to 0}j(\mathscr{E})$; 
i.e., the critical current $j_{\rm c}$ corresponds to the balance current at the minimum energy state. 
The critical current has been estimated from the linearized LLG equation as done in, for example, Ref. [\onlinecite{grollier03}]. 
Although the critical current is often regarded as the switching current in spin torque switching, 
$j_{\rm c}$ does not equal the switching current in the present system due to the following reasons. 
First, the initial state of the magnetization in the present system is a higher energy state than $\mathscr{E}_{\rm min}$. 
This is because microwaves applied from $t=0$ change the shape of the energy landscape for $t \ge 0$, 
and therefore, the initial state, $\mathbf{m}(0)=+\mathbf{e}_{z}$, is no longer a minimum energy state for $t \ge 0$, as mentioned above. 
Thus, an instability of the initial state does not relate to $j_{\rm c}$ in the present case. 
Second, the magnitude of $j_{\rm c}$ is sometimes smaller than that of $j(\mathscr{E})$ ($\mathscr{E}>\mathscr{E}_{\rm min}$). 
In such case, $j_{\rm c}$ does not satisfy Eq. (\ref{eq:jsw_def}), and therefore, cannot be regarded as a switching current. 
An example can be found in Sec. \ref{sec:Region 3} below. 
We emphasize that the instability of the equilibrium state does not guarantee the switching, 
and therefore, the critical current $j_{\rm c}$ and the switching current are not necessarily same. 

% ===================================================================================================================================================================================== %

The definition of the switching current, Eq. (\ref{eq:jsw_def}), with Eq. (\ref{eq:jE}) 
is based on the assumption that the work done by spin torque during a precession on a constant energy curve, 
Eq. (\ref{eq:W_s}), is finite. 
This assumption is valid in the present case. 
The switching current is proportional to the damping constant 
because the switching occurs as a result of the competition between the spin and damping torques. 
On the other hand, there are other spin torque switching problems 
where Eq. (\ref{eq:jsw_def}) cannot be used to evaluate the switching current. 
An example is the spin torque switching by spin Hall effect excited by a direct current [\onlinecite{dyakonov71,hirsch99,kato04,maekawa06}]
with a direct magnetic field applied in the direction of the electric current. 
In this case, the averaged LLG equation is no longer applicable to estimate the switching current 
because the averaged work done by spin torque becomes zero, and therefore, 
a steady precession on a constant energy curve cannot be excited. 
The switching in this case occurs as a result of the competition between the precession and spin torques, 
and thus, the switching current becomes independent of the damping constant [\onlinecite{liu12,lee13}]. 

% ===================================================================================================================================================================================== %

It is preferable to derive analytical formulas of the balance current $j(\mathscr{E})$, 
which will be helpful to explicitly apprehend the relation between the material parameters and the switching current. 
To derive the explicit form of $j(\mathscr{E})$, 
the analytical formulas of $\mathscr{W}_{\rm s}$ and $\mathscr{W}_{\alpha}$ should be derived. 
In principle, $\mathscr{W}_{\rm s}$ and $\mathscr{W}_{\alpha}$ are obtained by substituting 
the solution of $\mathbf{m}^{\prime}$ precessing on a constant energy curve of $\mathscr{E}$ into the integrands; 
i.e., the solution of $d \mathbf{m}^{\prime}/dt=-\gamma \mathbf{m}^{\prime} \times \bm{\mathcal{B}}$ is necessary. 
However, the equation $d \mathbf{m}^{\prime}/dt=-\gamma \mathbf{m}^{\prime} \times \bm{\mathcal{B}}$ is usually a nonlinear equation for two variables, 
and therefore, it is difficult to obtain the solution of $\mathbf{m}^{\prime}$. 
Several works, nevertheless, have derived the analytical formulas of $j(\mathscr{E})$ in the other systems 
[\onlinecite{bertotti04,serpico05,hillebrands06,newhall13,taniguchi13PRB1,taniguchi13PRB2,taniguchi13,pinna13,pinna14,taniguchi15}]. 
For example, Ref. [\onlinecite{taniguchi13PRB1}] derived the analytical formulas of $\mathscr{W}_{\rm s}$ and $\mathscr{W}_{\alpha}$ 
in a ferromagnetic multilayer, in which both the free and pinned layers are in-plane magnetized. 
Reference [\onlinecite{taniguchi13}] derived $\mathscr{W}_{\rm s}$ and $\mathscr{W}_{\alpha}$ of a spin torque oscillator 
consisting of a perpendicularly magnetized free layer and an in-plane magnetized pinned layer. 
These analytical results can be obtained because the system has high symmetry. 
Unfortunately however, it is difficult to derive the analytical formula of Eq. (\ref{eq:jE}) in the present system for general $\mathscr{E}$. 
Both $H_{\rm ac}$ and $(2\pi f/\gamma)$ terms in Eq. (\ref{eq:B_energy}) 
act as external magnetic fields along the $x^{\prime}$ and $z^{\prime}$ directions, respectively. 
Then, the total external field points to a tilted direction in the $x^{\prime}z^{\prime}$-plane, 
and breaks the symmetry along the $z^{\prime}$ axis. 
Therefore, the solution of $d \mathbf{m}^{\prime}/dt=-\gamma \mathbf{m}^{\prime} \times \bm{\mathcal{B}}$, 
as well as the analytical formulas of $\mathscr{W}_{\rm s}$ and $\mathscr{W}_{\alpha}$ for an arbitrary $\mathscr{E}$, can be hardly obtained. 
Therefore, $j(\mathscr{E})$ in the present work was evaluated numerically [\onlinecite{taniguchi14PRB}] in most cases; 
see Appendix \ref{sec:AppendixA}. 
It might be however possible to calculate $j(\mathscr{E})$ analytically in a particular case; 
see Appendix \ref{sec:AppendixB}. 

% ===================================================================================================================================================================================== %

\subsection{Region 1}
\label{sec:Region 1}

\begin{figure}%[p]
\centerline{\includegraphics[width=1.0\columnwidth]{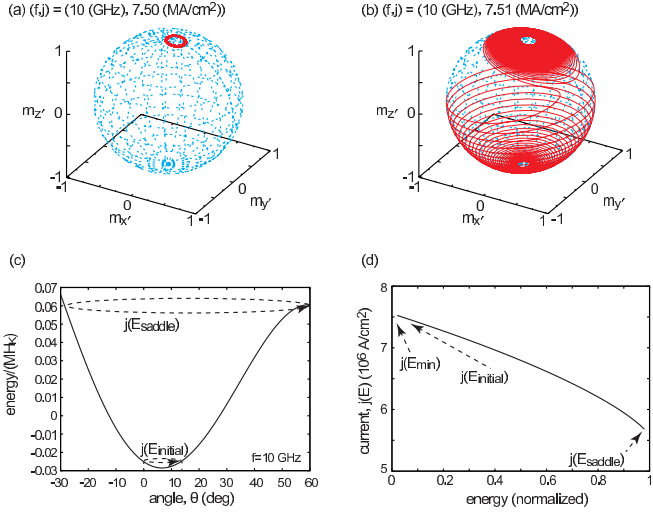}}%\vspace{-3.0ex}
\caption{
        Typical magnetization dynamics in region 1 near the switching current. 
        The microwave frequency is $10$ GHz, while the current density is (a) $7.50$ and (b) $7.51 \times 10^{6}$ A/cm${}^{2}$. 
        (c) The energy landscape near the initial state and the saddle point.
        (d) The relation between the energy $\mathscr{E}$ and the balance current $j(\mathscr{E})$ 
            from the metastable state to the saddle point. 
            The horizontal axis is normalized as $(\mathscr{E}-\mathscr{E}_{\rm min})/(\mathscr{E}_{\rm saddle}-\mathscr{E}_{\rm min})$, 
            where $\mathscr{E}_{\rm min}$ and $\mathscr{E}_{\rm saddle}$ are the values of $\mathscr{E}$ 
            at the metastable state and the saddle point, respectively. 
         \vspace{-3ex}}
\label{fig:fig6}
\end{figure}

Here, we discuss the switching current in region 1. 
Figures \ref{fig:fig6}(a) and \ref{fig:fig6}(b) show typical magnetization dynamics near the switching current. 
The microwave frequency is $f=10$ GHz, and the current is (a) $7.50$ and (b) $7.51 \times 10^{6}$ A/cm${}^{2}$. 
As shown, the magnetization precesses many times during the switching. 
This fact guarantees the above assumption to define the balance current. 
Now, let us consider the relation between the balance current and the switching current. 

% ===================================================================================================================================================================================== %

% ===================================================================================================================================================================================== %

% ===================================================================================================================================================================================== %

% ===================================================================================================================================================================================== %

Figure \ref{fig:fig6}(c) schematically shows the energy landscape for $f=10$ GHz. 
The range of $\mathscr{E}$ shown in this figure is 
$\mathscr{E}_{\rm min} < \mathscr{E} < \mathscr{E}_{\rm saddle}$, 
where $\mathscr{E}_{\rm min}$ and $\mathscr{E}_{\rm saddle}$ are the values of $\mathscr{E}$ 
at the metastable state and the saddle point, respectively. 
As emphasized above, 
because of the presence of the microwave field $H_{\rm ac}$ along the $x^{\prime}$ direction, 
the metastable state shifts from the $z^{\prime}$ axis 
with the angle $\theta \simeq 6.6^{\circ}$. 
Remember that the initial state, $\mathbf{m}(0)=+\mathbf{e}_{z}$ or equivalently $\theta=0$, is 
energetically higher than the metastable state, 
i.e., the energy density at the initial state, $\mathscr{E}_{\rm initial}=\mathscr{E}(\mathbf{m}=+\mathbf{e}_{z})$, 
is larger than $\mathscr{E}_{\rm min}$. 
Note also that the initial state is located at a lower energy state than the saddle point, 
and therefore, the magnetization must overcome the saddle point to switch its direction. 

% ===================================================================================================================================================================================== %

Figure \ref{fig:fig6}(d) shows 
the dependence of the balance current $j(\mathscr{E})$ on the energy density $\mathscr{E}$. 
The left and right ends of $j(\mathscr{E})$ correspond to $j(\mathscr{E}_{\rm min})$ and $j(\mathscr{E}_{\rm saddle})$, respectively, 
while $j(\mathscr{E}_{\rm initial})$ is located in between them. 
As shown, $j(\mathscr{E})$ monotonically decreases with increasing $\mathscr{E}$. 
The physical meaning of Fig. \ref{fig:fig6}(d) is as follows. 
Starting from zero current, 
when the current value is less than $j(\mathscr{E}_{\rm initial})$, 
the damping overcomes the spin torque, 
and the magnetization relaxes to the metastable state. 
When the current reaches $j(\mathscr{E}_{\rm initial})$, 
the spin torque just cancels the damping torque, 
and the magnetization precesses on the constant energy curve including the initial state. 
For a current larger than $j(\mathscr{E}_{\rm initial})$, 
the spin torque overcomes the damping torque, 
and thus, the magnetization moves from the initial state 
to a higher energy state, $\mathscr{E}>\mathscr{E}_{\rm initial}$. 
According to Fig. \ref{fig:fig6}(d), 
the key point for the initiation of the magnetization movement toward a higher energy state 
is to have $j(\mathscr{E}_{\rm initial})$ larger than $j(\mathscr{E})$ at a higher energy state. 
Then, the spin torque and the damping torque do not balance on any constant energy curve of $\mathscr{E}(>\mathscr{E}_{\rm initial})$, 
and the magnetization overcomes the saddle point by climbing up the energy landscape. 
Therefore, the switching current in this case is given by 
\begin{equation}
  j_{\rm sw}
  =
  j(\mathscr{E}_{\rm initial}),
  \label{eq:jsw_1}
\end{equation}
according to Eq. (\ref{eq:jsw_def}). 
The switching current estimated from Eq. (\ref{eq:jsw_1}) is $7.50 \times 10^{6}$ A/cm${}^{2}$, 
showing a good agreement with the numerical result. 
Note that both the damping torque ($-\alpha \gamma \mathbf{m}^{\prime}\times(\mathbf{m}^{\prime}\times\bm{\mathcal{B}})$) 
and the torque due to microwaves ($\alpha 2\pi f \mathbf{m}^{\prime}\times(\mathbf{e}_{z^{\prime}}\times\mathbf{m}^{\prime})$) 
in Eq. (\ref{eq:LLG_rotating_frame}) prevent the switching. 
The spin torque should overcome these torques to switch the magnetization. 
Therefore, the switching current is relatively large in this region. 

% ===================================================================================================================================================================================== %

% ===================================================================================================================================================================================== %

\begin{figure}%[p]
\centerline{\includegraphics[width=0.5\columnwidth]{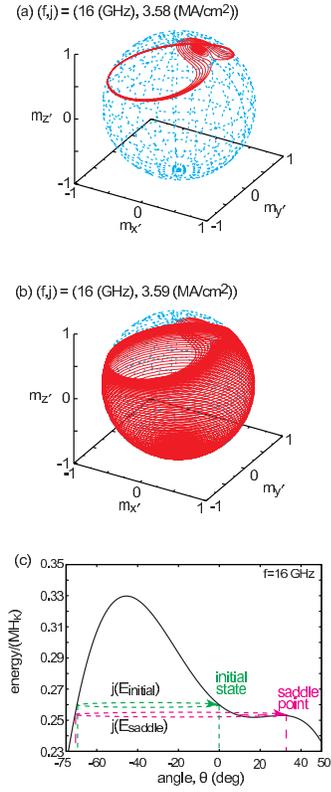}}%\vspace{-3.0ex}
\caption{
        Typical magnetization dynamics in region 2 near the switching current, 
        where the current densities are (a) $3.58$ and (b) $3.59 \times 10^{6}$ A/cm${}^{2}$, respectively. 
        The microwave frequency is $f=16$ GHz. 
        (c) The energy landscape of $\mathscr{E}$ for $f=16$ GHz. 
            The integral ranges of $j(\mathscr{E})$ in Eq. (\ref{eq:jE}), for $j(\mathscr{E}_{\rm initial})$ and $j(\mathscr{E}_{\rm saddle})$, are 
            schematically shown. 
         \vspace{-3ex}}
\label{fig:fig7}
\end{figure}

% ===================================================================================================================================================================================== %

% ===================================================================================================================================================================================== %

% ===================================================================================================================================================================================== %

\subsection{Region 2}
\label{sec:Region 2}

The switching current in region 2 is estimated as follows.  
Figures \ref{fig:fig7}(a) and \ref{fig:fig7}(b) show 
the magnetization dynamics at $f=16$ GHz, 
where the current densities are (a) $3.58$ and (b) $3.59 \times 10^{6}$ A/cm${}^{2}$, respectively. 
Contrary to the case of region 1 shown in Fig. \ref{fig:fig6}(a), 
the trajectory of the magnetization shows a relatively large amplitude precession around the $z^{\prime}$ axis. 
Figure \ref{fig:fig7}(c) schematically shows such precession on the energy landscape of $\mathscr{E}$. 
Remember that the initial state of the magnetization in region 2 
is located at a higher energy state than the saddle point. 
Thus, the precession trajectory in the energy landscape goes around the mountain of the potential around the unstable state. 
This phenomenon is contrary to the case of region 1 
in which the precession trajectory is inside the valley of the potential around the metastable state. 

% ===================================================================================================================================================================================== %

The calculation of the balance current requires an attention in this case. 
For example, let us consider the calculation of $j(\mathscr{E}_{\rm saddle})$. 
As shown in Fig. \ref{fig:fig7}(c), 
there are three angles, $\theta \simeq -70.7^{\circ}$, $9.8^{\circ}$, and $30.5^{\circ}$, having the same energy with $\mathscr{E}_{\rm saddle}$, 
where $\theta \simeq 30.5^{\circ}$ corresponds to the real saddle point. 
To calculate Eq. (\ref{eq:jE}), the precession period of the precession between 
$\theta\simeq -70.7^{\circ}$ and $30.5^{\circ}$ should be chosen as the boundaries of the integrals 
because a curve connecting these points corresponds to the precession trajectory shown in Fig. \ref{fig:fig7}(a). 

% ===================================================================================================================================================================================== %

Two balance currents characterize this system. 
The first one is the balance current including the initial state, $j(\mathscr{E}_{\rm initial})$, 
and the second one is the balance current including the saddle point, $j(\mathscr{E}_{\rm saddle})$, 
where both precession trajectories are shown in Fig. \ref{fig:fig7}(c). 
The evaluated values of $j(\mathscr{E}_{\rm initial})$ and $j(\mathscr{E}_{\rm saddle})$ are 
$3.23$ and $3.55 \times 10^{6}$ A/cm${}^{2}$, respectively. 
This means that, when a current larger than $j(\mathscr{E}_{\rm initial})$ and smaller than $j(\mathscr{E}_{\rm saddle})$ is applied, 
the initial state is destabilized. 
Then, the magnetization starts to move from the initial state to the saddle point 
along the gradient of the energy landscape in Fig. \ref{fig:fig7}(c). 
However, the magnetization cannot switch its direction 
because such current is still insufficient to cross the saddle point. 
%Therefore, the magnetization falls into the meta-stable state. 
On the other hand, when the current is larger than $j(\mathscr{E}_{\rm saddle})$, 
the magnetization can cross the saddle point, 
and switch its direction. 
Therefore, the switching current in the region 2 is related to the balance current via 
\begin{equation}
  j_{\rm sw}
  =
  j(\mathscr{E}_{\rm saddle}).
  \label{eq:jsw_2}
\end{equation}
When the magnetization is on the constant energy curve of $\mathscr{E}_{\rm saddle}$ in Fig. \ref{fig:fig7}(c), 
both the spin and damping torques point to the switched state, 
while the torque due to microwaves ($\alpha 2\pi f \mathbf{m}^{\prime}\times(\mathbf{e}_{z^{\prime}}\times\mathbf{m}^{\prime})$) 
in Eq. (\ref{eq:LLG_rotating_frame}) still prevents the switching. 
Therefore, the switching current becomes relatively small compared with the case in region 1. 

% ===================================================================================================================================================================================== %

% ===================================================================================================================================================================================== %

\begin{figure}%[p]
\centerline{\includegraphics[width=1.0\columnwidth]{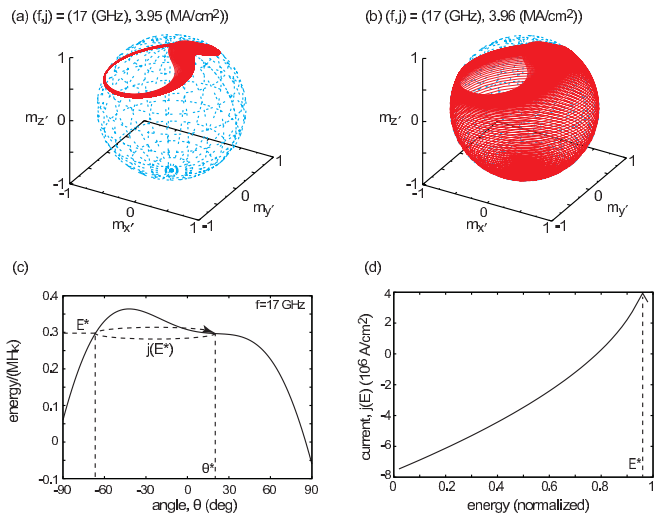}}%\vspace{-3.0ex}
\caption{
        Typical magnetization dynamics in region 3 near the switching current. 
        The microwave frequency is $17$ GHz, while the current density is (a) $3.95$ and (b) $3.96 \times 10^{6}$ A/cm${}^{2}$. 
        (c) The energy landscape for $f=17$ GHz. 
        (d) The relation between the energy $\mathscr{E}$ and the balance current $j(\mathscr{E})$ 
            from the metastable state to the saddle point. 
            The horizontal axis is normalized as $(\mathscr{E}-\mathscr{E}_{\rm min})/(\mathscr{E}_{\rm max}-\mathscr{E}_{\rm min})$, 
            where $\mathscr{E}_{\rm min}$ and $\mathscr{E}_{\rm max}$ are the values of $\mathscr{E}$ 
            at the stable state and the unstable point, respectively. 
         \vspace{-3ex}}
\label{fig:fig8}
\end{figure}

% ===================================================================================================================================================================================== %

% ===================================================================================================================================================================================== %

% ===================================================================================================================================================================================== %

\subsection{Region 3}
\label{sec:Region 3}

Here, let us discuss the switching in region 3. 
Figures \ref{fig:fig8}(a) and \ref{fig:fig8}(b) show typical magnetization dynamics 
around the switching current, where $f=17$ GHz. 
Similarly to the case of region 2, 
the trajectory saturates to a certain orbit when the current is slightly smaller than the switching current, 
as shown in Fig. \ref{fig:fig8}(a). 
However, let us remind the reader that the energy landscape of $\mathscr{E}$ in region 3 
does not have a saddle point, 
contrary to the case of region 2. 
Figure \ref{fig:fig8}(c) shows the energy landscape of $\mathscr{E}$ for $f=17$ GHz. 
We notice that, starting from the initial state, $\theta=0$, to the switched state, 
the gradient of $\mathscr{E}$ once decreases until $\theta \simeq 21.6^{\circ}$, 
and then increases again above $\theta \simeq 21.6^{\circ}$. 
In other words, the potential $\mathscr{E}$ has a point $\theta^{*}$ 
at which $\partial^{2}\mathscr{E}/\partial \theta^{2}|_{(\theta,\varphi)=(\theta^{*},0)}=0$. 
We denote the value of $\mathscr{E}$ corresponding to $(\theta,\varphi)=(\theta^{*},0)$, as $\mathscr{E}^{*}$. 

Figure \ref{fig:fig8}(d) shows the dependence of the balance current on the energy $\mathscr{E}$. 
The minimum energy corresponds to the stable (switched) state ($\theta \simeq 180^{\circ}$), 
while the maximum energy corresponds to the unstable state located near $\theta\simeq -42.2^{\circ}$ in Fig. \ref{fig:fig8}(c). 
Although the potential has only one minimum, 
a finite current is necessary to switch the magnetization 
because the second term on the right-hand side of Eq. (\ref{eq:LLG_rotating_frame}), 
$\alpha 2\pi f \mathbf{m}^{\prime} \times (\mathbf{e}_{z^{\prime}} \times \mathbf{m}^{\prime})$, 
prevents the switching, 
as in the case of microwave-assisted magnetization reversal [\onlinecite{taniguchi14PRB}]. 
The sign of $j(\mathscr{E})$ changes with decreasing the energy 
because the direction of the damping torque changes for $m_{z^{\prime}} \gtrsim 0$ and $m_{z^{\prime}} \lesssim 0$. 

We find that the balance current $j(\mathscr{E})$ in $\mathscr{E}_{\rm min} < \mathscr{E} < \mathscr{E}_{\rm max}$ 
has a maximum at $\mathscr{E}=\mathscr{E}^{*}$, as shown in Fig. \ref{fig:fig8}(d).  
%where $\mathscr{E}_{\rm min}$ and $\mathscr{E}_{\rm max}$ are the values of $\mathscr{E}$ 
%corresponding to the stable (switched) state and the unstable state, respectively. 
The physical meaning of this result is as follows. 
Starting from zero current, 
when the current becomes slightly larger than $j(\mathscr{E}_{\rm initial})$, 
the magnetization is destabilized by the spin torque; 
i.e., the spin torque overcomes the damping torque. 
Then, the magnetization starts to move in the direction of the switched state. 
When the current is larger than $j(\mathscr{E}_{\rm initial})$ and smaller than $j(\mathscr{E}^{*})$, 
the magnetization shows a steady precession on a constant energy curve 
corresponding to an energy between $\mathscr{E}_{\rm initial}$ and $\mathscr{E}^{*}$. 
On the other hand, when the current becomes larger than $j(\mathscr{E}^{*})$, 
the spin torque does not balance with the damping torque on any constant energy curve of $\mathscr{E}$. 
Therefore, the magnetization abandons the steady precession, and instead, moves to the switched state.  
Therefore, the switching current in region 3 is given by 
\begin{equation}
  j_{\rm sw}
  =
  j(\mathscr{E}^{*}).
  \label{eq:jsw_3}
\end{equation}
The numerically evaluated value of Eq. (\ref{eq:jsw_3}) is $3.95 \times 10^{6}$ A/cm${}^{2}$, 
showing a good agreement with the results in Figs. \ref{fig:fig8}(a) and \ref{fig:fig8}(b). 

% ===================================================================================================================================================================================== %

% ===================================================================================================================================================================================== %

\begin{figure}%[p]
\centerline{\includegraphics[width=1.0\columnwidth]{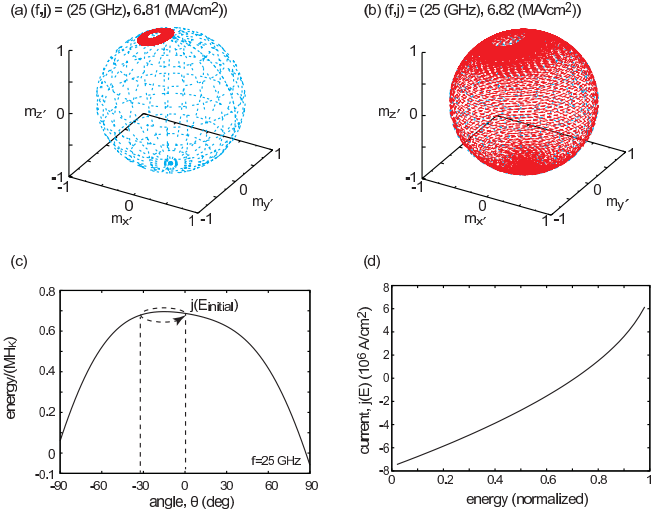}}%\vspace{-3.0ex}
\caption{
        Typical magnetization dynamics in region 4 near the switching current. 
        The microwave frequency is $25$ GHz while the current density is (a) $6.81$ and (b) $6.82 \times 10^{6}$ A/cm${}^{2}$. 
        (c) The energy landscape for $f=25$ GHz. 
        (d) The relation between the energy $\mathscr{E}$ and the balance current $j(\mathscr{E})$ 
            from the metastable state to the saddle point. 
            The horizontal axis is normalized as $(\mathscr{E}-\mathscr{E}_{\rm min})/(\mathscr{E}_{\rm max}-\mathscr{E}_{\rm min})$, 
            where $\mathscr{E}_{\rm min}$ and $\mathscr{E}_{\rm max}$ are the values of $\mathscr{E}$ 
            at the stable state and the unstable point, respectively. 
         \vspace{-3ex}}
\label{fig:fig9}
\end{figure}

% ===================================================================================================================================================================================== %

% ===================================================================================================================================================================================== %

% ===================================================================================================================================================================================== %

\subsection{Region 4}
\label{sec:Region 4}

The switching current in region 4 is estimated as follows. 
Figures \ref{fig:fig9}(a) and \ref{fig:fig9}(b) show typical magnetization dynamics, 
where the microwave frequency is $f=25$ GHz. 
The magnetization precesses many times during the switching, 
guaranteeing the validity to use the balance current to estimate the switching current. 
The energy landscape of $\mathscr{E}$ in this region has only one stable state, 
and the gradient of $\mathscr{E}$ becomes monotonic from the initial state to the stable state, 
as shown in Fig. \ref{fig:fig9}(c). 
A typical dependence of the balance current on $\mathscr{E}$ is shown in Fig. \ref{fig:fig9}(d). 
As shown, $j(\mathscr{E})$ decreases with decreasing $\mathscr{E}$. 
This means that, once the initial state is destabilized, 
the spin torque does not balance with the damping torque, 
and moves to the switched state. 
Therefore, according to Eq. (\ref{eq:jsw_def}), 
the switching current in region 4 is given by 
\begin{equation}
  j_{\rm sw}
  =
  j(\mathscr{E}_{\rm initial}).
  \label{eq:jsw_4}
\end{equation}
The evaluated value of Eq. (\ref{eq:jsw_4}) for $f=25$ GHz is $6.82 \times 10^{6}$ A/cm${}^{2}$, 
which is in good agreement with the numerical simulation shown in Figs. \ref{fig:fig9}(a) and \ref{fig:fig9}(b).

% ===================================================================================================================================================================================== %

% ===================================================================================================================================================================================== %

\begin{figure}%[p]
\centerline{\includegraphics[width=1.0\columnwidth]{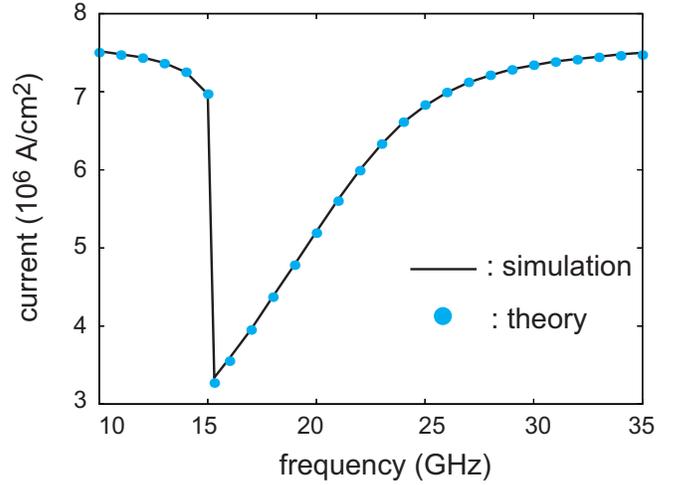}}%\vspace{-3.0ex}
\caption{
         Relation between the microwave frequency and the switching current. 
         The solid line is obtained from the numerical simulation of the LLG equation, Eq. (\ref{eq:LLG}), 
         while the dots are obtained by the balance currents, Eqs. (\ref{eq:jsw_1})-(\ref{eq:jsw_4}). 
         \vspace{-3ex}}
\label{fig:fig10}
\end{figure}

% ===================================================================================================================================================================================== %

% ===================================================================================================================================================================================== %

% ===================================================================================================================================================================================== %

\subsection{Summary of analyses}
\label{sec:Summary of analyses}

In Fig. \ref{fig:fig10}, 
we summarize the comparison of the switching currents 
obtained from the numerical simulation of Eq. (\ref{eq:LLG}) (solid line) 
and those estimated from Eqs. (\ref{eq:jsw_1})-(\ref{eq:jsw_4}). 
As shown, the theoretical formulas, Eqs. (\ref{eq:jsw_1})-(\ref{eq:jsw_4}), 
show good agreement with the numerical results, indicating the validity of our analysis. 

The above analyses of the switching currents for the four regions reveal the reason why 
the switching current is minimized at a certain frequency. 
When the microwave frequency is low (region 1), 
the magnetization initially stays a lower energetic state than the saddle point. 
The switching current is relatively large because the spin torque should bring the magnetization to the saddle point 
by overcoming the damping torque and the torque due to the microwaves, 
$\alpha 2\pi f \mathbf{m}\times(\mathbf{e}_{z^{\prime}}\times \mathbf{m})$ in Eq. (\ref{eq:LLG_rotating_frame}). 
On the other hand, when the microwave frequency becomes close to the optimized frequency (region 2), 
the magnetization initially locates at a higher energetic state than the saddle point. 
In this case, when the magnetization comes to a large-amplitude trajectory as shown in Fig. \ref{fig:fig7}(a), 
the spin and damping torques move the magnetization to the switched state, 
although the torque due to the microwaves still prevents the switching. 
Therefore, the switching current becomes suddenly low in region 2. 
The torque due to the microwaves preventing the reversal becomes large with increasing the frequency 
because the magnitude of this torque is proportional to the microwave frequency. 
Therefore, the switching current increases with increasing the frequency in regions 3 and 4, 
even though the potential $\mathscr{E}$ has only one minimum. 
Regarding the derivations of Eqs. (\ref{eq:jsw_1})-(\ref{eq:jsw_4}), 
the present results depend on the choice of the initial state. 
The theoretical approach based on the balance current, however, will be applicable 
even when a different initial state is chosen. 

% ===================================================================================================================================================================================== %

% ===================================================================================================================================================================================== %

\begin{figure}%[p]
\centerline{\includegraphics[width=1.0\columnwidth]{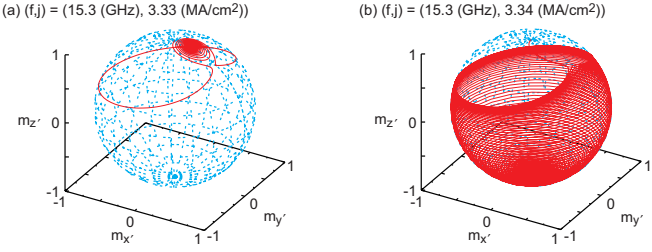}}%\vspace{-3.0ex}
\caption{
         Magnetization dynamics at the optimized frequency, $f=15.3$ GHz, 
         where the current density is (a) $3.33$ and (b) $3.34 \times 10^{6}$ A/cm${}^{2}$. 
         \vspace{-3ex}}
\label{fig:fig11}
\end{figure}

% ===================================================================================================================================================================================== %

% ===================================================================================================================================================================================== %

\section{Optimized frequency and minimized switching current}
\label{sec:Optimized frequency and minimized switching current}

In this section, we discuss the relation between the optimized frequency, the minimized switching current, and the physical parameters. 
The switching current is minimized 
around the boundary between region 1 and region 2. 
Remember that the definition of region 1 was that 
the initial state of the magnetization is located at a lower energy state than 
the saddle point. 
Therefore, the frequency determining 
the boundary between region 1 and region 2 
can be found by the relation that 
the energies of the initial state and the saddle point are the same. 
Such frequency, $f_{\rm b}$, can be estimated from the following equations: 
\begin{equation}
\begin{split}
  &
  -H_{\rm ac}
  \sin \theta_{\rm i}
  +
  \frac{2\pi f_{\rm b}}{\gamma}
  \cos\theta_{\rm i}
  -
  \frac{H_{\rm K}}{2}
  \cos^{2}\theta_{\rm i}
\\
  &=
  -H_{\rm ac}
  \sin \theta_{\rm d}
  +
  \frac{2\pi f_{\rm b}}{\gamma}
  \cos\theta_{\rm d}
  -
  \frac{H_{\rm K}}{2}
  \cos^{2}\theta_{\rm d},
  \label{eq:minimum_frequency_condition1}
\end{split}
\end{equation}
\begin{equation}
\begin{split}
  &
  -H_{\rm ac}
  \cos\theta_{\rm d}
  -
  \frac{2\pi f_{\rm b}}{\gamma}
  \sin \theta_{\rm d}
  +
  H_{\rm K}
  \sin\theta_{\rm d}
  \cos\theta_{\rm d}
  =
  0,
  \label{eq:minimum_frequency_condition2}
\end{split}
\end{equation}
where $\theta_{\rm i}$ and $\theta_{\rm d}$ are the zenith angles $\theta$ 
corresponding to the initial state and the saddle point, respectively. 
Equation (\ref{eq:minimum_frequency_condition1}) expresses the relation that 
the energies at the initial state and the saddle point are the same. 
Equation (\ref{eq:minimum_frequency_condition2}) implies that 
the gradient of the energy is zero at the saddle point. 
Solving Eqs. (\ref{eq:minimum_frequency_condition1}) and (\ref{eq:minimum_frequency_condition2}) 
with respect to $f_{\rm b}$ and $\theta_{\rm d}$, 
we found that the frequency determining the boundary 
between region 1 and region 2 is given by 
\begin{equation}
\begin{split}
  f_{\rm b}
  &=
  \frac{\gamma H_{\rm K}}{2\pi}
  \frac{z(1+z)}{2}
\\
  &
  \simeq 
  \frac{\gamma H_{\rm K}}{2\pi}
  \left[
    1
    -
    3 
    \left(
      \frac{H_{\rm ac}}{2 H_{\rm K}}
    \right)^{2/3}
    +
    \left(
      \frac{H_{\rm ac}}{2 H_{\rm K}}
    \right)^{4/3}
  \right.
\\
  &\ \ \ \ 
  \left.
   +
   \frac{1}{3}
   \left(
     \frac{H_{\rm ac}}{2 H_{\rm K}}
   \right)^{2}
   +
   \cdots
 \right].
 \label{eq:minimum_frequency}
\end{split}
\end{equation}
Here $z=\cos\theta_{\rm d}$ determined from Eqs. (\ref{eq:minimum_frequency_condition1}) and (\ref{eq:minimum_frequency_condition2}), 
or equivalently $\theta_{\rm d}$, can be expressed in terms of $H_{\rm ac}/H_{\rm K}$ as 
\begin{equation}
\begin{split}
  z
  =&
  1
  -
  2
  \left(
    \frac{H_{\rm ac}}{2 H_{\rm K}}
  \right)^{2/3}
  -
  \frac{2}{3}
  \left(
    \frac{H_{\rm ac}}{2 H_{\rm K}}
  \right)^{4/3}
\\
  &
  -
  \frac{2}{3}
  \left(
    \frac{H_{\rm ac}}{2 H_{\rm K}}
  \right)^{2}
  +
  \cdots,
  \label{eq:saddle_point1}
\end{split}
\end{equation}
or 
\begin{equation}
  \theta_{\rm d}
  =
  2 
  \left(
    \frac{H_{\rm ac}}{2 H_{\rm K}}
  \right)^{1/3}
  +
  \frac{H_{\rm ac}}{3 H_{\rm K}}
  +
  \frac{28}{45}
  \left(
    \frac{H_{\rm ac}}{2 H_{\rm K}}
  \right)^{5/3}
  +
  \cdots.
  \label{eq:saddle_point2}
\end{equation}
The higher order terms of $H_{\rm ac}/H_{\rm K}$ are negligible for $H_{\rm ac}/H_{\rm K} \ll 1$. 

% ===================================================================================================================================================================================== %

Strictly speaking, the optimized frequency belongs to region 2. 
In fact, the magnetization dynamics at the optimized frequency, $f=15.3$ GHz, 
shown in Figs. \ref{fig:fig11}(a) and \ref{fig:fig11}(b), are similar to Figs. \ref{fig:fig8}(a) and \ref{fig:fig8}(b). 
Therefore, the optimized frequency is the frequency slightly larger than Eq. (\ref{eq:minimum_frequency}). 
We note that the evaluated value of Eq. (\ref{eq:minimum_frequency}) is $15.2$ GHz, 
which shows good agreement with the optimized frequency, $f=15.3$ GHz. 
We also note that this optimized frequency is smaller than the FMR frequency, $f_{\rm FMR}=\gamma H_{\rm K}/(2\pi) \simeq 21$ GHz. 
Note that the position of the metastable state in Fig. \ref{fig:fig7}(c) moves to a large-$\theta$ region 
and the height of the saddle point becomes relatively low for a large $H_{\rm ac}/H_{\rm K}$, 
which make it easy to set the initial state higher than the saddle point by a slow microwave frequency. 
Therefore, the optimized frequency decreases with increasing $H_{\rm ac}/H_{\rm K}$. 

% ===================================================================================================================================================================================== %

% ===================================================================================================================================================================================== %

\begin{figure*}%[p]
\centerline{\includegraphics[width=2.0\columnwidth]{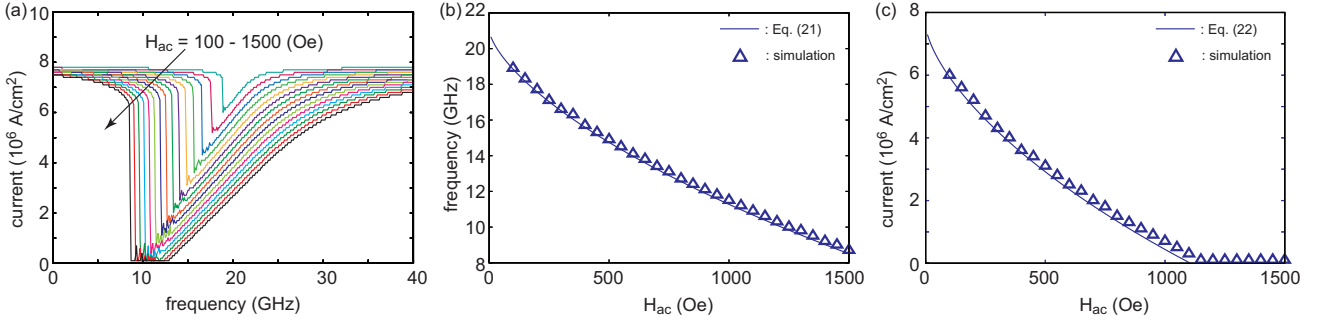}}%\vspace{-3.0ex}
\caption{
        (a) Numerically evaluated switching currents as a function of microwave frequency 
            for various $H_{\rm ac}=100-1500$ Oe, where the step of $H_{\rm ac}$ is 100 Oe. 
        (b) The microwave frequency minimizing the switching current obtained from (a) (solid line), 
            and the optimized frequency obtained from Eq. (\ref{eq:minimum_frequency}) (triangle). 
            The numerically evaluated switching current is sometimes minimized for a certain range of the frequency. 
            In such cases, the optimized frequency is defined as the minimum frequency in this range. 
        (c) The minimum switching current obtained from (a) (solid line), 
            and the same obtained from Eq. (\ref{eq:minimized_current}) (triangle). 
            The minimized switching current obtained from the numerical simulation is zero above $H_{\rm ac}=1150$ Oe, 
            indicating that the magnetization switching is achieved solely by the microwave when the microwave frequency is optimized. 
         \vspace{-3ex}}
\label{fig:fig12}
\end{figure*}

% ===================================================================================================================================================================================== %

% ===================================================================================================================================================================================== %

The analytical solution of the minimized switching current is also obtained in a similar way. 
The details of the calculation are summarized in Appendix \ref{sec:AppendixB}. 
The result is 
\begin{equation}
  j_{\rm sw}^{\rm min}
  =
  \frac{2 \alpha eMd}{\hbar \eta}
  \left(
    \frac{2\pi f_{\rm b}}{\gamma}
    +
    H_{\rm K}
    \frac{\mathscr{N}_{\alpha}}{\mathscr{N}_{\rm s}}
  \right),
  \label{eq:minimized_current}
\end{equation}
where $\mathscr{N}_{\rm s}$ and $\mathscr{N}_{\alpha}$ are given by 
\begin{equation}
\begin{split}
  \mathscr{N}_{\rm s}
  =&
  -\frac{\sqrt{1-z^{2}}(1-z)(1-z^{2})}{4}
\\
  & \times
  \left[
    2 
    \left(
      1
      +
      2z
      +
      3z^{2}
    \right)
    \tan^{-1}
    \sqrt{1+2z}
  \right.
\\
  &
  \left.
    \ \ \ \ +
    3 z 
    \sqrt{1 + 2z}
  \right],
  \label{eq:Is_sol}
\end{split}
\end{equation}
\begin{equation}
\begin{split}
  \mathscr{N}_{\alpha}
  =
  &
  \frac{\sqrt{1-z^{2}}(1-z)^{3}}{24}
\\
  &
  \times
  \left[
    \sqrt{1+2z}
    \left(
      8
      +
      20z
      +
      29z^{2}
      +
      15z^{3}
    \right)
  \right.
\\
  &
  \left.
   \ \ \ \ +
    6z
    \left(
      5
      +
      13z
      +
      13z^{2}
      +
      5z^{3}
    \right)
    \tan^{-1}
    \sqrt{1+2z}
  \right].
  \label{eq:Ialpha_sol}
\end{split}
\end{equation}
In the limit of the zero-microwave ($H_{\rm ac} \to 0$), 
we find that $f_{\rm b} \to \gamma H_{\rm K}/(2\pi)$, 
$\mathscr{N}_{\rm s} \to 0$, 
$\mathscr{N}_{\alpha} \to 0$, 
and $\mathscr{N}_{\alpha}/\mathscr{N}_{\rm s} \to 0$. 
Then, Eq. (\ref{eq:minimized_current}) reproduces the switching current solely by spin torque, Eq. (\ref{eq:jc_spin_torque}). 
The evaluated value of Eq. (\ref{eq:minimized_current}) is $3.22 \times 10^{6}$ A/cm${}^{2}$, 
showing a good agreement with the numerical result shown in Fig. \ref{fig:fig11}. 

% ===================================================================================================================================================================================== %

Notice here that both the optimized frequency, Eq. (\ref{eq:minimum_frequency}), 
and the minimized switching current, Eq. (\ref{eq:minimized_current}), are functions of $H_{\rm ac}/H_{\rm K}$. 
In other words, the ratio between the optimized frequency and the FMR frequency, 
as well as that between the minimized switching current and the switching current solely by spin torque, 
depends on $H_{\rm ac}/H_{\rm K}$ only. 
The above numerical and analytical results were obtained from $H_{\rm ac}=450$ Oe. 
In the following, we compare these analytical results with numerical results for various values of $H_{\rm ac}$ 
to confirm the validities of Eqs. (\ref{eq:minimum_frequency}) and (\ref{eq:minimized_current}). 
Figure \ref{fig:fig12}(a) shows the numerically evaluated switching currents as a function of the microwave frequency for various $H_{\rm ac}$. 
The optimized frequency and the minimized switching current obtained from these results are summarized by triangles 
in Figs. \ref{fig:fig12}(b) and \ref{fig:fig12}(c), respectively. 
In these figures, we also show analytical results obtained from Eqs. (\ref{eq:minimum_frequency}) and (\ref{eq:minimized_current}) by solid lines. 
The numerical and analytical results show good agreement in Fig. \ref{fig:fig12}(b). 
A good agreement between the numerical and analytical results is also obtained for the minimized switching current, as shown in Fig. \ref{fig:fig12}(c). 
These results guarantee the validity of Eqs. (\ref{eq:minimum_frequency}) and (\ref{eq:minimized_current}). 
Both the optimized frequency and minimized switching current are decreasing functions of $H_{\rm ac}/H_{\rm K}$. 
Note that the analytical current, Eq. (\ref{eq:minimized_current}), is zero near $H_{\rm ac}\simeq 1100$ Oe ($H_{\rm ac}/H_{\rm K} \simeq 0.15$), 
and becomes negative for $H_{\rm ac}>1100$ Oe. 
In our definition, the negative current means that the spin torque prevents the switching from $\mathbf{m}(0)=+\mathbf{e}_{z}$. 
Since Eq. (\ref{eq:minimized_current}) is obtained from the balance current, 
this result implies that the spin torque is unnecessary to switch the magnetization for $H_{\rm ac}>1100$ Oe. 
In fact, the numerically evaluated minimized switching current is zero above $H_{\rm ac}=1150$ Oe, 
indicating that the magnetization switching is achieved solely by the microwave when the microwave frequency is optimized. 
The magnetization switching solely by the microwave will be an interesting topic in the field of the microwave-assisted magnetization reversal. 
The topic is however not the main scope of this paper, 
and will be discussed briefly in Appendix \ref{sec:AppendixC}. 
We should also remind the reader that the above theory is applicable for a small damping $(\alpha \ll 1)$, as mentioned in Sec. \ref{sec:Balance current}. 

% ===================================================================================================================================================================================== %

% ===================================================================================================================================================================================== %

\section{Conclusion}
\label{sec:Conclusion}

In conclusion, 
we investigated the switching current of a perpendicular ferromagnet by spin transfer torque 
in the presence of a circularly polarized microwave both numerically and analytically. 
Numerical simulation of the LLG equation revealed that 
the switching current is significantly reduced 
when the microwave frequency is in a certain range. 
The switching current can become even zero when the microwave frequency is optimized 
and the amplitude of the microwaves becomes relatively high.
%For typical material parameters, the minimized switching current is less than a half of the switching current 
%which would be necessary when the switching is operated solely by the spin torque. 
We developed a theory to evaluate the switching current 
from the LLG equation averaged over a constant energy curve. 
It was found that the switching current should be classified into four regions, 
depending on the values of the microwave frequencies. 
Based on the analysis, we derived an analytical formula of the optimized frequency 
at which the switching current is minimized. 
The analytical formula of the minimized switching current is also obtained. 
These analytical formulas show good agreement with the numerical results 
for a wide range of the microwave field amplitude. 
The minimized switching current decreases with increasing the amplitude of the microwave field. 
The results provide a pathway to achieve both high thermal stability and low switching current simultaneously.

% ===================================================================================================================================================================================== %

\section*{Acknowledgement}

The authors are thankful to Takehiko Yorozu for his contributions 
on the derivation of the analytical formulas. 
The authors also appreciate Yoshishige Suzuki having valuable discussions. 
T. T. expresses gratitude to 
Shinji Yuasa, Kay Yakushiji, Koji Ando, Akio Fukushima, 
Takayuki Nozaki, Makoto Konoto, Hidekazu Saito, 
Satoshi Iba, Aurelie Spiesser, 
Yoichi Shiota, Sumito Tsunegi, Ryo Hiramatsu, 
Shinji Miwa, Takahide Kubota, Atsushi Sugihara, 
Hiroki Maehara, and Ai Emura for their support and encouragement. 
This work was supported by the NEDO Normally Off Computing project. 

% ===================================================================================================================================================================================== %

\appendix

% ===================================================================================================================================================================================== %

\section{Calculation procedures of Eqs. (\ref{eq:W_s}) and (\ref{eq:W_alpha})}
\label{sec:AppendixA}

The balance current can be evaluated 
by calculating the time integral in Eqs. (\ref{eq:W_s}) and (\ref{eq:W_alpha}). 
In principle, the solution $\mathbf{m}^{\prime}(t)$ of the LLG equation on a constant energy curve, 
$d \mathbf{m}^{\prime}/dt = -\gamma \mathbf{m}^{\prime} \times \bm{\mathcal{B}}$, 
is necessary to perform the time integral. 
Using the following technique, however, Eqs. (\ref{eq:W_s}) and (\ref{eq:W_alpha}) can be calculated 
without the time-dependent solution of $\mathbf{m}^{\prime}(t)$ obtained from Eq. (\ref{eq:LLG_rotating_frame}). 
Note that the integration variable can be transformed from the time $t$ to $m_{z^{\prime}}$ 
by using the $z$ component of the LLG equation on a constant energy curve, $d m_{z^{\prime}}/dt= \gamma H_{\rm ac} m_{y^{\prime}}$. 
In other words, $\oint dt$ in Eqs. (\ref{eq:W_s}) and (\ref{eq:W_alpha}) is replaced with 
$2 \int d m_{z^{\prime}}/(\gamma H_{\rm ac}m_{y^{\prime}})$. 
The numerical factor $2$ appears 
by restricting the integral range to $m_{y^{\prime}}>0$ and 
due to the symmetry of the system with respect to the $x^{\prime}z^{\prime}$ plane. 
Because the LLG equation conserves the magnetization magnitude, 
$m_{y^{\prime}}$ appearing in Eqs. (\ref{eq:W_s}) and (\ref{eq:W_alpha}) 
can be replaced by 
$\sqrt{1-m_{x^{\prime}}^{2}-m_{z^{\prime}}^{2}}$. 
Also, from Eq. (\ref{eq:B_energy}), 
$m_{x^{\prime}}$ can be expressed in terms of $m_{z^{\prime}}$ as 
\begin{equation}
  m_{x^{\prime}}
  =
  \frac{1}{H_{\rm ac}}
  \left(
    -\frac{\mathscr{E}}{M}
    +
    \frac{2\pi f}{\gamma}
    m_{z^{\prime}}
    -
    \frac{H_{\rm K}}{2}
    m_{z^{\prime}}^{2}
  \right).
  \label{eq:m_x}
\end{equation}
Therefore, the integrand in Eqs. (\ref{eq:W_s}) and (\ref{eq:W_alpha}) is expressed using $m_{z^{\prime}}$ only. 
The integral range can be determined 
from Eq. (\ref{eq:B_energy}) by fixing the value $\mathscr{E}$. 

% ===================================================================================================================================================================================== %

\section{Calculation of minimized switching current}
\label{sec:AppendixB}

In this Appendix, we show the details of the derivation of Eq. (\ref{eq:minimized_current}). 
As mentioned in Sec. \ref{sec:Balance current}, 
it is rather complex to derive the general solution of the balance current. 
In general, the balance current, in Eq. (\ref{eq:jE}), 
as well as in Eqs. (\ref{eq:W_s}) and (\ref{eq:W_alpha}), depends on 
$H_{\rm ac}/H_{\rm K}$ and $f/H_{\rm K}$ through $\mathscr{E}$, Eq. (\ref{eq:B_energy}). 
However, the minimized switching current can be expressed in terms of $H_{\rm ac}/H_{\rm K}$ only 
because $H_{\rm ac}$ and $f$ in this case are not independent of each other 
due to the conditions Eqs. (\ref{eq:minimum_frequency_condition1}) and (\ref{eq:minimum_frequency_condition2}). 
In fact, the frequency $f$ and the saddle point $z=\cos\theta_{\rm d}$ are expressed 
in terms of $H_{\rm ac}/H_{\rm K}$, as shown in Eqs. (\ref{eq:minimum_frequency}), (\ref{eq:saddle_point1}), and (\ref{eq:saddle_point2}). 
Since $H_{\rm ac}/H_{\rm K} \ll 1$ in typical experiments of microwave-assisted magnetization reversal, 
it is sufficient to express Eqs. (\ref{eq:W_s}) and (\ref{eq:W_alpha}) by lower orders of $H_{\rm ac}/H_{\rm K}$ 
for the evaluation of the minimized switching current. 

%====================================================================================================================================================== %

% ===================================================================================================================================================================================== %

% ===================================================================================================================================================================================== %

\begin{figure}%[p]
\centerline{\includegraphics[width=1.0\columnwidth]{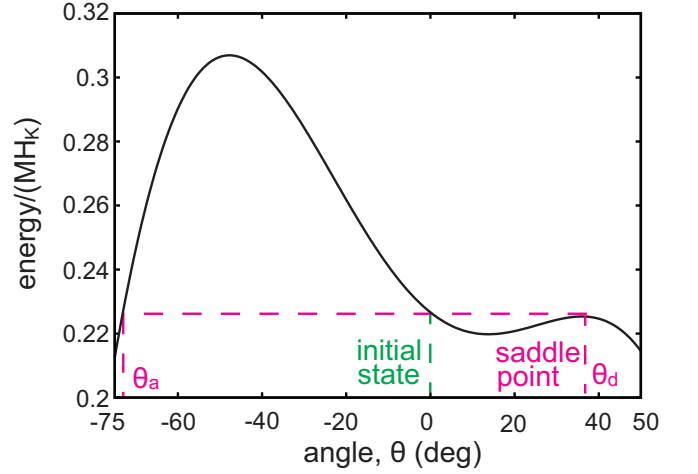}}%\vspace{-3.0ex}
\caption{
         The energy landscape for the optimized frequency. 
         The angle $\theta$ corresponding to the saddle point is denoted as $\theta_{\rm d}$, 
         while the angle corresponding to the same energy with the saddle point is denoted as $\theta_{\rm a}$. 
         \vspace{-3ex}}
\label{fig:fig13}
\end{figure}

% ===================================================================================================================================================================================== %

% ===================================================================================================================================================================================== %

As mentioned in Sec. \ref{sec:Optimized frequency and minimized switching current}, 
the optimized frequency belongs to region 2. 
Therefore, the integral ranges of Eqs. (\ref{eq:W_s}) and (\ref{eq:W_alpha}) are 
determined from the energy landscape shown in Fig. \ref{fig:fig13}, 
where the energies at the initial state and the saddle point are identical. 
The angle $\theta_{\rm d}$ corresponding to the saddle point is $\theta_{\rm d} \simeq 36.9^{\circ}$ for our parameters. 
With these parameters there is another point $\theta_{\rm a}$ having the same energy with 
the initial state ($\theta=0^{\circ}$) and the saddle point $\theta_{\rm d}$, 
which is $\theta_{\rm a} \simeq -73.7^{\circ}$. 
As mentioned in Appendix \ref{sec:AppendixA}, 
the time integrals in Eqs. (\ref{eq:W_s}) and (\ref{eq:W_alpha}) can be 
converted to the integral with respect to $m_{z^{\prime}}$. 
The integral range is then $x \le m_{z^{\prime}} \le z$, 
where $z$ is given by  Eq. (\ref{eq:saddle_point1}) while $x=\cos\theta_{\rm a}$. 
We notice that 
\begin{equation}
  \theta_{\rm a}
  =
  -2 \theta_{\rm d}. 
\end{equation}
This relation can be proved as follows. 
For simplicity, let us introduce the normalized field and frequency as $a=H_{\rm ac}/H_{\rm K}$ and $b=2\pi f/(\gamma H_{\rm K})$. 
Then, the normalized energy at the saddle point is 
\begin{equation}
  \varepsilon
  =
  -a u 
  +
  b z 
  -
  \frac{1}{2}
  z^{2},
  \label{eq:normalized_energy_saddle}
\end{equation}
where $\varepsilon=\mathscr{E}_{\rm saddle}/(MH_{\rm K})$ and $u=\sin\theta_{\rm d}$. 
Note that $u^{2}+z^{2}=1$, 
and Eq. (\ref{eq:minimum_frequency_condition2}) can be rewritten as $-az-bu+uz=0$, 
i.e., 
\begin{equation}
  b
  =
  z
  \left(
    1
    -
    \frac{a}{u}
  \right). 
  \label{eq:b_relation}
\end{equation}
Substituting these relations into Eq. (\ref{eq:normalized_energy_saddle}), 
we find that 
\begin{equation}
  a
  =
  \frac{u}{2}
  \left(
    z^{2}
    -
    2 \varepsilon
  \right). 
  \label{eq:a_relation}
\end{equation}
Note also that $x=\cos\theta_{\rm a}$ and $v=\sin\theta_{\rm a}$ 
satisfies $\varepsilon=-av + bx - x^{2}/2$, according to the definition of $\theta_{\rm a}$. 
Then, we find 
\begin{equation}
  v
  =
  \frac{2bx-x^{2}-2\varepsilon}{2a}.
  \label{eq:v_relation}
\end{equation}
These relations are independent of the choice of the initial state. 
Now let us assume that the energy at the saddle point equals that at the initial state, $\theta=0$. 
This means that $\varepsilon=b-1/2$. 
Then, Eqs. (\ref{eq:b_relation}) and (\ref{eq:a_relation}) are rewritten as 
\begin{equation}
  b
  =
  \frac{z(1+z)}{2},
  \label{eq:b_relation_new}
\end{equation}
\begin{equation}
  a
  =
  \frac{u(1-z)}{2}.
  \label{eq:a_relation_new}
\end{equation}
Therefore, the normalized energy at the saddle point becomes 
\begin{equation}
  \varepsilon
  =
  -\frac{1}{2}
  \left(
    1
    -
    z
    -
    z^{2}
  \right).
  \label{eq:normalized_energy}
\end{equation}
Using these relations and Eq. (\ref{eq:v_relation}), 
we find that 
\begin{equation}
  1
  -
  v^{2}
  -
  x^{2}
  =
  \frac{(x-z)^{2}(1-x)[x-(2z^{2}-1)]}{(1-z)(1-z^{2})}.
\end{equation}
According to the definition of $\theta_{\rm a}$, 
$1-v^{2}-x^{2}=0$. 
Therefore, $x$ should be one of $z$, $1$, and $2z^{2}-1$. 
The solutions of $x=z$ and $x=1$ are reasonable 
because the energy $\mathscr{E}$ at $\theta_{\rm a}$ equals to 
that at the saddle point ($m_{z^{\prime}}=z$) and the initial state ($m_{z^{\prime}}=1$). 
The other solution, $x=2z^{2}-1$, corresponds to $x=\cos\theta_{\rm a}$. 
This solution indicates that $\theta_{\rm a}=-2\theta_{\rm d}$ 
because the cosine function satisfies $\cos 2 \theta=2 \cos^{2}\theta-1$ and $\cos\theta=\cos(-\theta)$. 

%====================================================================================================================================================== %

Using Eqs. (\ref{eq:b_relation}), (\ref{eq:a_relation}), (\ref{eq:normalized_energy}), and $u^{2}+z^{2}=1$, 
we find that $m_{x^{\prime}}$ given by Eq. (\ref{eq:m_x}) and $m_{y^{\prime}}=\sqrt{1-m_{x^{\prime}}^{2}-m_{z^{\prime}}^{2}}$ are 
expressed in terms of $m_{z^{\prime}}$ and $z=\cos\theta_{\rm d}$ as 
\begin{equation}
  m_{x^{\prime}}
  =
  \frac{(1-z-z^{2})+z(1+z)m_{z^{\prime}}-m_{z^{\prime}}^{2}}{\sqrt{1-z^{2}} (1-z)},
  \label{eq:mx_def}
\end{equation}
\begin{equation}
  m_{y^{\prime}}
  =
  \frac{\sqrt{(1-m_{z^{\prime}}) (z-m_{z^{\prime}})^{2} (1+m_{z^{\prime}}-2z^{2})}}{\sqrt{1-z^{2}}(1-z)}.
  \label{eq:my_def}
\end{equation}
These results indicate that all quantities determining the minimized switching current can be 
expressed in terms of $H_{\rm ac}/H_{\rm K}$ through $z=\cos\theta_{\rm d}$. 

%====================================================================================================================================================== %

To obtain the analytical formula of the minimized switching current, 
let us define the following integrals from Eqs. (\ref{eq:W_s}) and (\ref{eq:W_alpha}): 
\begin{equation}
  \mathscr{N}_{\rm s}
  =
  \int 
  \frac{d m_{z^{\prime}}}{m_{y^{\prime}}}
  \left[
    \tilde{\bm{\mathcal{B}}}
    \cdot
    \mathbf{e}_{z^{\prime}}
    -
    \left(
      \mathbf{m}^{\prime}
      \cdot
      \mathbf{e}_{z^{\prime}}
    \right)
    \left(
      \mathbf{m}^{\prime}
      \cdot
      \tilde{\bm{\mathcal{B}}}
    \right)
  \right],
  \label{eq:Is_def}
\end{equation}
\begin{equation}
  \mathscr{N}_{\alpha}
  =
  \int 
  \frac{d m_{z^{\prime}}}{m_{y^{\prime}}}
  \left[
    \tilde{\bm{\mathcal{B}}}^{2}
    -
    \left(
      \mathbf{m}^{\prime}
      \cdot
      \tilde{\bm{\mathcal{B}}}
    \right)^{2}
  \right],
  \label{eq:Ialpha_def}
\end{equation}
where $\tilde{\bm{\mathcal{B}}}=\bm{\mathcal{B}}/H_{\rm K}=(a,0,-b+m_{z^{\prime}})$ is the normalized magnetic field in the rotating frame. 
Equations (\ref{eq:Is_def}) and (\ref{eq:Ialpha_def}) relate to 
Eqs. (\ref{eq:W_s}) and (\ref{eq:W_alpha}) via 
$\mathscr{W}_{\rm s}=2 M H_{\rm K} (\gamma H_{\rm s}-\alpha 2\pi f) \mathscr{N}_{\rm s}/(\gamma H_{\rm ac})$ and 
$\mathscr{W}_{\alpha}=-2 \alpha M H_{\rm K}^{2} \mathscr{N}_{\alpha}/H_{\rm ac}$, respectively. 
The minimized switching current is then given by Eq. (\ref{eq:minimized_current}), where 
\begin{equation}
  f_{\rm b}
  =
  \frac{\gamma H_{\rm K}b}{2\pi}, 
\end{equation}
is the optimized frequency given by Eq. (\ref{eq:minimum_frequency}). 
Using Eqs. (\ref{eq:b_relation_new}), (\ref{eq:a_relation_new}), (\ref{eq:mx_def}), and (\ref{eq:my_def}), 
the integrands of Eqs. (\ref{eq:Is_def}) and (\ref{eq:Ialpha_def}) can be expressed in terms of $m_{z^{\prime}}$ and $z$. 
Explicitly, these are given by 
\begin{equation}
  \mathscr{N}_{\rm s}
  =
  -\int \!\!
  d m_{z^{\prime}}
  \frac{\sqrt{1-z^{2}}(1-z)\sqrt{1-m_{z^{\prime}}}(1+m_{z^{\prime}}+z)}{2 \sqrt{1+m_{z^{\prime}}-2z^{2}}},
  \label{eq:Is_def_sub}
\end{equation}
\begin{equation}
  \mathscr{N}_{\alpha}
  =
  \int \!\!
  d m_{z^{\prime}}
  \frac{\sqrt{1-z^{2}}(1-z)(z-m_{z^{\prime}})[2(1+z)-(m_{z^{\prime}}+z)^{2}]}{4 \sqrt{(1-m_{z^{\prime}})(1+m_{z^{\prime}}-2z^{2})}}.
  \label{eq:Ialpha_def_sub}
\end{equation}
The integral region is $\cos \theta_{\rm a} \le m_{z^{\prime}} \le \cos \theta_{\rm d}$, as shown in Fig. \ref{fig:fig13}, 
which can be expressed in terms of $z$ as $[x,z]=[2z^{2}-1,z]$. 
To perform these integrals, it is convenient to introduce a new variable $s$ as 
$s=\sqrt{(1-m_{z^{\prime}})/(1+m_{z^{\prime}}-2z^{2})}$. 
The integral region then becomes $(\infty,\sqrt{1/(2z+1)}]$. 
Then, we find 
\begin{equation}
\begin{split}
  \mathscr{N}_{\rm s}
  =&
  \frac{(1-z^{2})^{3/2} (1-z)}{2 (1+s^{2})^{2}}
\\
  &
  \left\{
    -s[1+2z+3z^{2}+s^{2}(-1+2z+5z^{2})] 
  \right.
\\
  &
  \left.
    + 
    (1+s^{2})^{2}(1+2z+3z^{2})\tan^{-1}s
  \right\}
  \bigg|_{\infty}^{\sqrt{1/(2z+1)}},
\end{split}
\end{equation}
\begin{equation}
\begin{split}
  \mathscr{N}_{\alpha}
  =
  &
  \frac{(1-z^{2})^{3/2}(1-z)^{2}}{12(1+s^{2})^{3}}
\\
  &
  \left\{
    s
    \left[
      3
      \left(
        2
        +
        5z
        +
        8z^{2}
        +
        5z^{3}
      \right)
    \right.
  \right.
\\
  &
      +
      4s^{2}
      \left(
        5
        +
        11 z
        +
        16 z^{2}
        +
        10z^{3}
      \right)      
\\
  &
    \left.
      +
      3 s^{4}
      \left(
        2
        +
        7z
        +
        16z^{2}
        +
        11z^{3}
      \right)
    \right]
\\
  &
  \left.
    -
    3 (1+s^{2})^{3}
    z
    \left(
      5
      +
      8z
      +
      5z^{2}
    \right)
    \tan^{-1}s
  \right\}
  \bigg|_{\infty}^{\sqrt{1/(2z+1)}}.
\end{split}
\end{equation}
These become Eqs. (\ref{eq:Is_sol}) and (\ref{eq:Ialpha_sol}), respectively, 
by using a formula $\tan^{-1}(1/\xi)={\rm sgn}(\xi)(\pi/2)-\tan^{-1} \xi$. 
Using Eqs. (\ref{eq:Is_sol}) and (\ref{eq:Ialpha_sol}), we can confirm that 
$\lim_{H_{\rm ac} \to 0}\mathscr{W}_{\rm s}=0$ and $\lim_{H_{\rm ac} \to 0}\mathscr{W}_{\alpha}=0$. 
The physical meaning of these limits is as follows. 
In the absence of the microwave ($H_{\rm ac} \to 0$), 
the initial state, $\mathbf{m}(0)=+\mathbf{e}_{z}$, corresponds to the energetically stable state, 
i.e., the minimum of the potential $\mathscr{E}$. 
The constant energy curve in this case becomes just a point, and thus, 
both the work done by spin torque and the dissipation due to damping during a precession on the constant energy curve are zero. 
We note that the explicit forms of Eqs. (\ref{eq:Is_sol}) and (\ref{eq:Ialpha_sol}), as well as Eq. (\ref{eq:minimum_frequency}), 
depend on the choice of the initial state, according to Eqs. (\ref{eq:minimum_frequency_condition1}) and (\ref{eq:minimum_frequency_condition2}). 
The initial condition, $\theta_{\rm i}=0$, in our calculation is reasonable when both the current and microwaves are applied to the free layer from $t=0$ 
because the equilibrium for $t<0$ in this case corresponds to this state. 
On the other hand, if only microwaves are applied during a certain time without current, 
the magnetization will relax to the metastable state shown in Fig. \ref{fig:fig3}(b). 
In this case, modification will be necessary in the above formulation. 

%====================================================================================================================================================== %

In the field of microwave-assisted magnetization reversal, 
it has been known that the switching field shows a minimum at a certain microwave frequency, 
similar to the present study. 
The theoretical conditions determining the optimized frequency were already derived in our previous work [\onlinecite{taniguchi14PRB}]. 
We noticed, however, that we can derive another form of the optimized frequency 
which is mathematically identical to the previous result but easier to use 
by performing an analytical calculation in a similar manner to that discussed above. 
%Contrary to the present study, an analytical formula of the optimized frequency 
%for the microwave assisted magnetization reversal has not yet been derived. 
%We noticed, however, that it can be derived by performing an analytical calculation in similar manners as discussed above. 
Although the optimized frequency solely by the microwave-assisted magnetization reversal is not a main target in this paper, 
we briefly summarize the calculations in Appendix \ref{sec:AppendixD} for comparison. 

%====================================================================================================================================================== %

\section{Magnetization switching solely by microwave}
\label{sec:AppendixC}

As mentioned in Sec. \ref{sec:Optimized frequency and minimized switching current}, 
the magnetization switching occurs solely by microwave when the microwave amplitude becomes relatively large. 
The microwave amplitude to switch the magnetization solely by the microwaves is obtained 
from the condition that the minimized switching current, Eq. (\ref{eq:minimized_current}), is negative. 
Therefore, the minimum amplitude of such microwaves is estimated from the equation, 
\begin{equation}
  \frac{2\pi f_{\rm b}}{\gamma}
  +
  H_{\rm K}
  \frac{\mathscr{N}_{\alpha}}{\mathscr{N}_{\rm s}}
  =
  0.
  \label{eq:condition_microwave_switch}
\end{equation}
The condition Eq. (\ref{eq:condition_microwave_switch}) depends on $H_{\rm ac}/H_{\rm K}$ only 
because $f_{\rm b}$, $\mathscr{N}_{\rm s}$, and $\mathscr{N}_{\alpha}$ are functions of $H_{\rm ac}/H_{\rm K}$; 
see Eq. (\ref{eq:minimum_frequency}), (\ref{eq:Is_sol}), (\ref{eq:Ialpha_sol}). 
Therefore, the condition that $H_{\rm ac}/H_{\rm K}>0.15$ to switch the magnetization solely by microwaves 
found in Sec. \ref{sec:Optimized frequency and minimized switching current} is independent of the choice of material parameters. 
Note that the condition depends on the initial state, as mentioned in Appendix \ref{sec:AppendixB}. 
Also another switching condition on the microwave field, $H_{\rm ac}/H_{\rm K}>\alpha/2$, derived from the LLG equation [\onlinecite{sun06a}], 
should also be satisfied, which is independent of the initial state but depends on the damping constant $\alpha$. 
Let us remind the reader that our theory using the averaged LLG equation is applicable for a small $\alpha$. 
In this case, the latter condition, $H_{\rm ac}/H_{\rm K}>\alpha/2$, is usually satisfied when $H_{\rm ac}/H_{\rm K}>0.15$ is satisfied. 

%====================================================================================================================================================== %

\section{Optimized frequency of microwave-assisted magnetization reversal}
\label{sec:AppendixD}

In this Appendix, let us show an analytical formula of the optimized frequency for microwave-assisted magnetization reversal. 
Note that the theoretical conditions determining the optimized frequency were already derived 
in our previous work [\onlinecite{taniguchi14PRB}]. 
Here, we derive the formula mathematically identical to the previous result 
but easier to use for analyzing the microwave-assisted magnetization reversal. 
In this section, %we add the notation "MAMR" to quantities related to the microwave assisted magnetization reversal (MAMR), 
we add the prime marks to quantities related to the microwave-assisted magnetization reversal, 
such as the effective energy density $\mathscr{E}$, its saddle point $z$, 
and the integrals $\mathscr{W}_{\rm s}$ and $\mathscr{W}_{\alpha}$, 
to distinguish from those quantities used in the main text and Appendix \ref{sec:AppendixB}. 
Note that the prime marks in $\mathbf{m}^{\prime}=(m_{x^{\prime}},m_{y^{\prime}},m_{z^{\prime}})$ are 
used to emphasize the fact that they are the magnetization components in the rotating frame. 
In the microwave-assisted magnetization reversal, a direct field $H$ is applied to the negative $z$ direction, 
and the spin torque is absent. 
Thus, the effective energy density is given by (see also Ref. [\onlinecite{taniguchi14PRB}]) 
\begin{equation}
\begin{split}
  \mathscr{E}^{\prime}
  =&
  -MH_{\rm ac}
  m_{x^{\prime}}
  +
  M
  \left(
    H
    +
    \frac{2\pi f}{\gamma}
  \right)
  m_{z^{\prime}}
\\
  &-
  \frac{MH_{\rm K}}{2}
  m_{z^{\prime}}^{2}.
  \label{eq:energy_MAMR}
\end{split}
\end{equation}
As investigated in Ref. [\onlinecite{taniguchi14PRB}], 
the optimized frequency, which was designated as the jump frequency in this paper, 
satisfies the conditions 
\begin{equation}
%  \frac{\partial \mathscr{E}_{\rm MAMR}}{\partial \theta}
  \frac{\partial \mathscr{E}^{\prime}}{\partial \theta}
  \bigg|_{\varphi=0}
  =
%  \frac{\partial^{2} \mathscr{E}_{\rm MAMR}}{\partial \theta^{2}}
  \frac{\partial^{2} \mathscr{E}^{\prime}}{\partial \theta^{2}}
  \bigg|_{\varphi=0}
  =
  0,
  \label{eq:MAMR_condition1}
\end{equation}
\begin{equation}
%  \mathscr{W}_{\rm s}^{\rm MAMR}(\mathscr{E}_{\rm MAMR,\ saddle})
%  +
%  \mathscr{W}_{\alpha}^{\rm MAMR}(\mathscr{E}_{\rm MAMR,\ saddle})
  \mathscr{W}_{\rm s}^{\prime}(\mathscr{E}_{\rm saddle}^{\prime})
  +
  \mathscr{W}_{\alpha}^{\prime}(\mathscr{E}_{\rm saddle}^{\prime})
  =
  0,
  \label{eq:MAMR_condition2}
\end{equation}
%where $\mathscr{W}_{\rm s}^{\rm MAMR}$ and $\mathscr{W}_{\alpha}^{\rm MAMR}$ are obtained 
where $\mathscr{W}_{\rm s}^{\prime}$ and $\mathscr{W}_{\alpha}^{\prime}$ are obtained 
by adding the external field $-H \mathbf{e}_{z^{\prime}}$ to $\bm{\mathcal{B}}$ in Eq. (\ref{eq:B_field}) 
and setting $H_{\rm s}=0$. 
The saddle point energy of Eq. (\ref{eq:energy_MAMR}) is denoted as $\mathscr{E}_{\rm saddle}^{\prime}$. %$\mathscr{E}_{\rm MAMR,\ saddle}$. 
Equations (\ref{eq:MAMR_condition1}) and (\ref{eq:MAMR_condition2}) are the theoretical conditions 
determining the optimized frequency of microwave-assisted magnetization reversal. 
Then, let us derive the explicit form of the frequency satisfying these conditions. 
From Eq. (\ref{eq:MAMR_condition1}), 
we notice that the angle %$\theta_{\rm MAMR,\ d}$ 
$\theta_{\rm d}^{\prime}$ corresponding to the saddle point satisfies the relation 
\begin{equation}
%  \sin\theta_{\rm MAMR,\ d}
  \sin\theta_{\rm d}^{\prime}
  =
  \left(
    \frac{H_{\rm ac}}{H_{\rm K}}
  \right)^{1/3}. 
\end{equation}
Also, the angle $\theta_{\rm a}^{\prime}$ satisfying %$\theta_{\rm MAMR,\ a}$ satisfying 
%$\mathscr{E}_{\rm MAMR}(\theta_{\rm MAMR,\ a})=\mathscr{E}_{\rm MAMR,\ saddle}$ and $\theta_{\rm MAMR,\ a} \neq \theta_{\rm MAMR,\ d}$ 
$\mathscr{E}^{\prime}(\theta_{\rm a}^{\prime})=\mathscr{E}_{\rm saddle}^{\prime}$ and $\theta_{\rm a}^{\prime} \neq \theta_{\rm d}^{\prime}$ 
is given by 
\begin{equation}
%  \theta_{\rm MAMR,\ a}
%  =
%  -3\theta_{\rm MAMR,\ d}. 
  \theta_{\rm a}^{\prime}
  =
  -3\theta_{\rm d}^{\prime}. 
\end{equation}
From these angles, we define %$z=\cos\theta_{\rm MAMR,\ d}$ and $x=\cos 3\theta_{\rm MAMR,\ d}=4z^{3}-3z$. 
\begin{equation}
  z^{\prime}
  =
  \cos
  \theta_{\rm d}^{\prime},
\end{equation}
and $x^{\prime}=\cos 3\theta_{\rm d}^{\prime}=4z^{\prime 3}-3z^{\prime}$. 
Similarly to Eq. (\ref{eq:normalized_energy}), 
the dimensionless energy at the saddle point is given by 
\begin{equation}
%  \varepsilon_{\rm MAMR}
  \varepsilon^{\prime}
  =
  -1
  +
  \frac{3}{2}
  z^{\prime 2}.
%  z^{2}.
\end{equation}
Also, $m_{x^{\prime}}$ and $m_{y^{\prime}}$ can be expressed in terms of %$z=\cos\theta_{\rm MAMR,\ d}$ and $m_{z^{\prime}}$ as 
$z^{\prime}=\cos\theta_{\rm d}^{\prime}$ and $m_{z^{\prime}}$ as 
\begin{equation}
  m_{x^{\prime}}
  =
  \frac{2-3z^{\prime 2}+2z^{\prime 2}m_{z^{\prime}}-m_{z^{\prime}}^{2}}{2 (1-z^{\prime 2})^{3/2}},
%  \frac{2-3z^{2}+2z^{2}m_{z^{\prime}}-m_{z^{\prime}}^{2}}{2 (1-z^{2})^{3/2}},
\end{equation}
\begin{equation}
  m_{y^{\prime}}
  =
  \frac{(z^{\prime}-m_{z^{\prime}}) \sqrt{(z^{\prime}-m_{z^{\prime}}) (m_{z^{\prime}}+3z^{\prime}-4z^{\prime 3})}}{2 (1-z^{\prime 2})^{3/2}}.
%  \frac{(z-m_{z^{\prime}}) \sqrt{(z-m_{z^{\prime}}) (m_{z^{\prime}}+3z-4z^{3})}}{2 (1-z^{2})^{3/2}}.
\end{equation}
Using these relations, we find that, similarly to Eqs. (\ref{eq:Is_def_sub}) and (\ref{eq:Ialpha_def_sub}), 
%$\mathscr{W}_{\rm s}^{\rm MAMR}$ and $\mathscr{W}_{\alpha}^{\rm MAMR}$ 
$\mathscr{W}_{\rm s}^{\prime}$ and $\mathscr{W}_{\alpha}^{\prime}$ 
determining the optimized frequency for the microwave-assisted magnetization reversal 
are given by 
%$\mathscr{W}_{\rm s}^{\rm MAMR}=-2\alpha M (H_{\rm K}^{2}/H_{\rm ac}) \mathscr{N}_{\rm s}^{\rm MAMR}$ 
%and $\mathscr{W}_{\alpha}^{\rm MAMR}=-2\alpha M (H_{\rm K}^{2}/H_{\rm ac}) \mathscr{N}_{\alpha}^{\rm MAMR}$, where 
$\mathscr{W}_{\rm s}^{\prime}=-2\alpha M (H_{\rm K}^{2}/H_{\rm ac}) \mathscr{N}_{\rm s}^{\prime}$ 
and $\mathscr{W}_{\alpha}^{\prime}=-2\alpha M (H_{\rm K}^{2}/H_{\rm ac}) \mathscr{N}_{\alpha}^{\prime}$, where 
%\begin{equation}
%  \mathscr{N}_{\rm s}^{\rm MAMR}
%  =
%  -\frac{2\pi f}{\gamma H_{\rm K}}
%  \int_{x}^{z} 
%  d m_{z^{\prime}}
%  \frac{(1-z^{2})^{3/2} (m_{z^{\prime}}+2z) \sqrt{z-m_{z^{\prime}}}}{\sqrt{m_{z^{\prime}}-4z^{3}+3z}},
%\end{equation}
%\begin{equation}
%  \mathscr{N}_{\alpha}^{\rm MAMR}
%  =
%  \int_{x}^{z}
%  d m_{z^{\prime}}
%  \frac{(1-z^{2})^{3/2}(m_{z^{\prime}}+3z)(z-m_{z^{\prime}})^{3/2}}{\sqrt{m_{z^{\prime}}-4z^{3}+3z}}.
%\end{equation}
\begin{equation}
  \mathscr{N}_{\rm s}^{\prime}
  =
  -\frac{2\pi f}{\gamma H_{\rm K}}
  \int_{x^{\prime}}^{z^{\prime}} 
  d m_{z^{\prime}}
  \frac{(1-z^{\prime 2})^{3/2} (m_{z^{\prime}}+2z^{\prime}) \sqrt{z^{\prime}-m_{z^{\prime}}}}{\sqrt{m_{z^{\prime}}-4z^{\prime 3}+3z^{\prime}}},
\end{equation}
\begin{equation}
  \mathscr{N}_{\alpha}^{\prime}
  =
  \int_{x^{\prime}}^{z^{\prime}}
  d m_{z^{\prime}}
  \frac{(1-z^{\prime 2})^{3/2}(m_{z^{\prime}}+3z^{\prime})(z^{\prime}-m_{z^{\prime}})^{3/2}}{\sqrt{m_{z^{\prime}}-4z^{\prime 3}+3z^{\prime}}}.
\end{equation}
Performing the integral, we find that 
%\begin{equation}
%  \mathscr{N}_{\rm s}^{\rm MAMR}
%  =
%  -\frac{12\pi^{2}f z^{4} \left(1-z^{2} \right)^{5/2}}{\gamma H_{\rm K}},
%\end{equation}
%\begin{equation}
%  \mathscr{N}_{\alpha}^{\rm MAMR}
%  = 
%  2 \pi 
%  z^{3}(1-z^{2})^{7/2}
%  \left(
%    1
%    +
%    5z^{2}
%  \right). 
%\end{equation}
\begin{equation}
  \mathscr{N}_{\rm s}^{\prime}
  =
  -\frac{12\pi^{2}f z^{\prime 4} \left(1-z^{\prime 2} \right)^{5/2}}{\gamma H_{\rm K}},
\end{equation}
\begin{equation}
  \mathscr{N}_{\alpha}^{\prime}
  = 
  2 \pi 
  z^{\prime 3}(1-z^{\prime 2})^{7/2}
  \left(
    1
    +
    5z^{\prime 2}
  \right). 
\end{equation}
Therefore, the optimized frequency for microwave-assisted magnetization reversal, 
which satisfies Eq. (\ref{eq:MAMR_condition2}) is given by 
%\begin{equation}
%\begin{split}
%  f_{\rm MAMR}
%  &=
%  \frac{\gamma H_{\rm K}}{6\pi z}
%  \left(
%    1
%    -
%    z^{2}
%  \right)
%  \left(
%    1
%    +
%    5z^{2}
%  \right)
%\\
%  &=
%  \frac{\gamma H_{\rm K}}{2\pi}
%  \frac{(H_{\rm ac}/H_{\rm K})^{2/3}}{\sqrt{1-(H_{\rm ac}/H_{\rm K})^{2/3}}}
%  \left[
%    2
%    -
%    \frac{5}{3}
%    \left(
%      \frac{H_{\rm ac}}{H_{\rm K}}
%    \right)^{2/3}
%  \right].
%  \label{eq:optimized_frequency_MAMR}
%\end{split}
%\end{equation}
\begin{equation}
\begin{split}
  f_{\rm MAMR}
  &=
  \frac{\gamma H_{\rm K}}{6\pi z^{\prime}}
  \left(
    1
    -
    z^{\prime 2}
  \right)
  \left(
    1
    +
    5z^{\prime 2}
  \right)
\\
  &=
  \frac{\gamma H_{\rm K}}{2\pi}
  \frac{(H_{\rm ac}/H_{\rm K})^{2/3}}{\sqrt{1-(H_{\rm ac}/H_{\rm K})^{2/3}}}
  \left[
    2
    -
    \frac{5}{3}
    \left(
      \frac{H_{\rm ac}}{H_{\rm K}}
    \right)^{2/3}
  \right].
  \label{eq:optimized_frequency_MAMR}
\end{split}
\end{equation}
This is the optimized (jump) frequency minimizing the switching field in the microwave-assisted magnetization reversal, 
which was formulated in our previous work [\onlinecite{taniguchi14PRB}] but was not derived explicitly. 
The validity of the formula was already confirmed in Ref. [\onlinecite{taniguchi14PRB}]. 
%in Ref. [\onlinecite{taniguchi14PRB}] but was not derived analytically. 
We notice that Eq. (\ref{eq:minimum_frequency}) is a decreasing function of $H_{\rm ac}/H_{\rm K}$, 
while Eq. (\ref{eq:optimized_frequency_MAMR}) is its increasing function. 
The reason is as follows.
According to Ref. [\onlinecite{taniguchi14PRB}], 
the switching below the optimized frequency occurs when the energy at the initial state is 
larger than that at the saddle point. 
This condition can be satisfied for a wide range of the microwave frequency when $H_{\rm ac}/H_{\rm K}$ 
because the saddle point energy becomes relatively low for a large $H_{\rm ac}/H_{\rm K}$. 
Therefore, the optimized frequency in microwave-assisted magnetization reversal increases with increasing $H_{\rm ac}/H_{\rm K}$. 
Equation (\ref{eq:optimized_frequency_MAMR}) can be either larger or smaller than the FMR frequency, $f_{\rm FMR}=\gamma H_{\rm K}/(2\pi)$, 
depending on the value of $H_{\rm ac}/H_{\rm K}$, 
contrary to Eq. (\ref{eq:minimum_frequency}), which is always smaller than the FMR frequency. 
The value of Eq. (\ref{eq:optimized_frequency_MAMR}) becomes $f_{\rm FMR}$ when $H_{\rm ac}/H_{\rm K} \simeq 0.52$. 

%====================================================================================================================================================== %

%====================================================================================================================================================== %

%\bibliography{biblist}

%====================================================================================================================================================== %

\end{document}